\documentclass[aps,prb,reprint,superscriptaddress]{revtex4-2}

\pdfoutput=1

\usepackage[utf8]{inputenc}
\usepackage{amsmath,amssymb,amsfonts,graphicx,verbatim,hyperref, mathtools, bbold, float}
\usepackage{physics}
\usepackage{soul, color}
\usepackage{algorithm, algorithmic}
\hypersetup{
	colorlinks=true,
	linkcolor=red,
	citecolor=blue,
	filecolor=blue,
}

\soulregister\ref{7}  
\soulregister\cite{7} 

\newcommand{\J}{\mathcal{J}}
\newcommand*\diff{\mathop{}\!\mathrm{d}}

\newcommand*{\im}{\, \mathrm{Im}}
\newcommand*{\re}{\, \mathrm{Re}}

\renewcommand*{\eqref}[1]{Eq.~(\ref{#1})}

\newcommand*{\figref}[1]{Fig.~\ref{#1}}

\newcommand*{\secref}[1]{Sec.~\ref{#1}}

\newcommand*{\appref}[1]{Appendix~\ref{#1}}

\newcommand{\hc}{\mathrm{h.c.}}

\begin{document}

\title{Near-Equilibrium Approach to Transport in Complex Sachdev-Ye-Kitaev Models}
\date{\today}
\author{Cristian \surname{Zanoci}}
\email{czanoci@mit.edu}
\affiliation{Department of Physics, Massachusetts Institute of Technology, 77 Massachusetts Avenue, Cambridge, Massachusetts 02139, USA}
\author{Brian \surname{Swingle}}
\email{bswingle@brandeis.edu}
\affiliation{Department of Physics, Brandeis University, Waltham, Massachusetts 02453, USA}

\begin{abstract}
We study the non-equilibrium dynamics of a one-dimensional complex Sachdev-Ye-Kitaev chain by directly solving for the steady state Green's functions in terms of small perturbations around their equilibrium values. The model exhibits strange metal behavior without quasiparticles and features diffusive propagation of both energy and charge. We explore the thermoelectric transport properties of this system by imposing uniform temperature and chemical potential gradients. We then expand the conserved charges and their associated currents to leading order in these gradients, which we can compute numerically and analytically for different parameter regimes. This allows us to extract the full temperature and chemical potential dependence of the transport coefficients. In particular, we uncover that the diffusivity matrix takes on a simple form in various limits and leads to simplified Einstein relations. At low temperatures, we also recover a previously known result for the Wiedemann-Franz ratio. Furthermore, we establish a relationship between diffusion and quantum chaos by showing that the diffusivity eigenvalues are upper bounded by the chaos propagation rate at all temperatures. Our work showcases an important example of an analytically tractable calculation of transport properties in a strongly interacting quantum system and reveals a more general purpose method for addressing strongly coupled transport.

\end{abstract}

\maketitle

\section{Introduction}
\label{sec:intro}

The study of quantum systems out of equilibrium can shed light on many outstanding questions regarding thermalization, transport, and quantum many-body chaos in condensed matter theories. In addition to these conceptual problems, there are many practical questions motivated by recent experimental advances in ultracold atoms and solid state systems, which present new avenues for investigating the non-equilibrium dynamics of many-body systems. In particular, quantum transport has garnered a lot of attention recently in an attempt to uncover new features of the dynamical processes governing the behavior of strongly interacting systems out of equilibrium. Despite numerous efforts, practical calculations of transport coefficients in quantum many-body systems remain challenging from both a theoretical and technical standpoint~\cite{bertini2021}, especially at low temperatures. 

One-dimensional models have emerged as prototypical examples for studying transport phenomena, partly due to their computational tractability. They usually consist of interacting particles or spins on a lattice that are driven away from equilibrium by certain external biases. The system then relaxes to a steady state dictated by its microscopic dynamics, which typically involves the transport of conserved quantities according to local conservation laws~\cite{zotos1997transport,kapustin2021}. These conserved charges and their associated currents are the quantities of interest.

A common implementation of this idea involves connecting the system to baths that drive it towards a desired steady state, where many transport properties are easily available~\cite{bertini2021,weimer2021,landi2021}. However, reaching this non-equilibrium steady state (NESS) in the hydrodynamic limit can be practically challenging~\cite{zanoci2021}, since most numerical techniques are usually limited to small systems and short evolution times. If we could bypass simulating the open-system non-equilibrium dynamics entirely and instead access the emergent NESS directly, we would be able to immediately find all the transport properties of the system. 

For a general class of models, we have previously shown that the local Green's functions in NESS are only slightly perturbed from their equilibrium values in the case of weak driving~\cite{zanoci2022}. This allowed us to find these non-equilibrium corrections explicitly in terms of the equilibrium Green's functions, without having to solve for the open-system dynamics. Our method is equivalent to a first order expansion in the local gradients, and thus falls under the umbrella of linear response theory. In this approximation, the temperature and chemical potential differences across the system are assumed to be small compared to their average values. Conveniently, most experimental setups studying transport in many-body systems also operate in the linear-response regime. 

The class of models in question consists of lattices built from the Sachdev-Ye-Kitaev (SYK) model~\cite{georges2000,georges2001,sachdev1993,parcollet1999,sachdev2015,kitaev2015simple,maldacena2016,kitaev2018,sarosi2017,rosenhaus2019,Gu2020,chowdhury2021sachdev} describing fermions with random all-to-all $q$-body interactions. In this paper, we will focus specifically on the complex fermion version of SYK~\cite{sachdev1993,parcollet1999,georges2000,georges2001,sachdev2015,Fu2016,davison2017,Bulycheva2017,Gu2020,tikhanovskaya2021a,tikhanovskaya2021b,chowdhury2021sachdev}, which has an additional conserved global $U(1)$ charge. This model displays a multitude of remarkable properties, ranging from an emergent approximate conformal symmetry at low temperatures~\cite{kitaev2015simple,sachdev2015,maldacena2016} to maximal many-body chaos~\cite{maldacena2016chaos}. In fact, the SYK model is holographically dual to extremal charged black holes with AdS$_2$ horizons~\cite{sachdev2010holographic,sachdev2010strange,sachdev2015,kitaev2015simple,maldacena2016,almheiri2015,maldacena2016ads,kitaev2018,engelsy2016}, and has a residual entropy directly connected to the Bekenstein-Hawking entropy of these black holes~\cite{sachdev2010holographic,sachdev2010strange,sachdev2015}. The model and its many variations~\cite{gu2017diffusion,cmjian2017,song2017,patel2018,chowdhury2018,guo2020,guo2019} belong to a class of systems realizing holographic quantum matter without quasiparticles, an thus represent a valuable platform for studying non-Fermi liquid behavior~\cite{sachdev1993,parcollet1999,sachdev2015,chowdhury2021sachdev}. Given the interesting physical properties of the SYK family of models, several experimental implementations have been recently proposed~\cite{franz2018,rahmani2019,danshita2017,wei2021,pikulin2017,chew2017,yang2018,chen2018,garcia2017,luo2019,babbush2019,behrends2022}.

The non-equilibrium dynamics of SYK models has been previously studied through various quench protocols~\cite{eberlein2017,bhattacharya2019,kuhlenkamp2020,samui2021,louw2021thermalization} or through couplings to external baths~\cite{almheiri2019,zhang2019,zanoci2022,cheipesh2020,haldar2020,chen2017,can2019} and Lindblad operators~\cite{sa2021lindbladian,kulkarni2021syk}. In particular, several questions pertaining to transport and chaos in higher-dimensional lattices of coupled SYK clusters have been addressed~\cite{gu2017diffusion,davison2017,song2017,cmjian2017,patel2018,chowdhury2018,guo2019,guo2020,zanoci2022}. These include many indicative properties of strange metals, such as diffusive propagation of energy~\cite{gu2017diffusion,davison2017,song2017} and resistivity that scales linearly with temperature~\cite{song2017,guo2020}. Moreover, it was shown that the same time-reparametrization field is responsible for the propagation of both low-energy modes and quantum chaos~\cite{gu2017diffusion,davison2017}, thus leading to a connection between energy diffusion and the butterfly velocity~\cite{hartnoll2014,blake2016_1,blake2016_2,blake2017,hartman2017,gu2017,chen2020,choi2021,hartnoll2021planckian,Blake2018,Blake2021}.

Nonetheless, the problem of characterizing transport for arbitrary model parameters remains mostly unsolved. Many of the previous approaches relied on the large $q$ limit or the low-temperature Schwarzian effective action to describe the energy and charge fluctuations~\cite{gu2017diffusion,gu2017,cmjian2017,choi2021}. These methods have a limited range of applicability and often do not lead to explicit solutions for the transport coefficients. In this paper, we propose a more general approach based on the expansion of the SYK Green's functions in the near-equilibrium regime, in the presence of constant temperature and chemical potential gradients. This allows us to compute the charge and energy currents throughout an SYK chain, and hence determine the associated diffusivities and conductivities. This method has the immediate advantage of delivering numerical results for any values of temperature, chemical potential, and $q$. Additionally, we obtain closed-form solutions in the limits of large $q$ and $q=2$.  

This work represents a natural extension of our previous analysis of energy diffusion in Majorana SYK models~\cite{zanoci2022}. Since complex fermions feature both charge and energy conservation, we were able to fully characterize the combined thermoelectric response of a strange metal and expose some of its most fascinating aspects. First, we studied the interplay of transport with various thermodynamic quantities and the physics of phase transitions. Second, we showed that the diffusivity matrix takes on a particular form at both high and low temperatures, as well as in the large $q$ limit. Third, we verified that the Wiedemann-Franz ratio approaches a known constant at zero temperature~\cite{davison2017}. Last, but not least, we related the eigenvalues of the diffusivity matrix to an upper bound set by chaos $D_\pm\leq v_B^2/\lambda_L$ at all temperatures and saturated in the conformal limit~\cite{gu2017diffusion,davison2017,choi2021,zanoci2022}. Our results provide concrete values for the transport coefficients that can be measured in the aforementioned experiments. But most importantly, they suggest a promising path towards more general studies of out-of-equilibrium phenomena in strongly interacting systems in which one directly accesses non-equilibrium steady states of interest.

The rest of the paper is structured as follows. In \secref{sec:model} we introduce our one-dimensional SYK model. Subsequently, in \secref{sec:methods} we describe in detail our approach to studying the equilibrium, non-equilibrium, and chaotic properties of this model. In \secref{sec:results} we review the phase diagram of the complex SYK model and present our main results  for the transport coefficients as a function of temperature and chemical potential. We also discuss a chaos bound on diffusivities in that section. We then provide a brief discussion of our findings and comment on possible extensions in \secref{sec:discussion}. The details of our calculations are available in the Appendix. 

\section{Model}
\label{sec:model}

The building block of our model is a complex SYK cluster~\cite{sachdev1993,parcollet1999,georges2000,georges2001,sachdev2015,Fu2016,davison2017,Bulycheva2017,Gu2020,tikhanovskaya2021a,tikhanovskaya2021b,chowdhury2021sachdev} with random all-to-all $q-$body interactions among $N$ fermions in $(0+1)$ dimensions. In order to study transport in this model, we generalize it to an infinitely-long one-dimensional chain (see \figref{fig:fig1}), where each site $x$ is an SYK cluster characterized by the Hamiltonian 

\begin{equation}
    H_0^x = \sum_{\{i\},\{j\}}J_{i_1\ldots i_{\frac{q}{2}}j_1\ldots j_{\frac{q}{2}}}^{(0)} (c_{i_1}^x)^\dagger\cdots (c_{i_{\frac{q}{2}}}^x)^\dagger c_{j_1}^x\cdots c_{j_{\frac{q}{2}}}^x, 
\end{equation}
where $q$ is an even integer and $\{i\}$ denotes the restricted sum over $1\leq i_1 < \cdots < i_{\frac{q}{2}}\leq N$. The neighboring sites of the chain interact via a similar Hamiltonian

\begin{equation}
    H_1^{x,x+1} = \sum_{\{i\},\{j\}}J_{i_1\ldots j_{\frac{q}{2}}}^{(1)} (c_{i_1}^{x})^\dagger\cdots (c_{i_{\frac{q}{2}}}^x)^\dagger c_{j_1}^{x+1}\cdots c_{j_{\frac{q}{2}}}^{x+1}+\hc.
\end{equation}

The fermions obey the standard anti-commutation relations $\{(c_i^x)^\dagger, c_j^{x'}\}=\delta_{ij}\delta_{xx'}$. Note that we choose the interaction term to consist of the same number of fermion operators from each site and one can consider adding more general interactions that may change the transport properties of the model~\cite{davison2017}. The SYK couplings are complex, independent Gaussian random variables with zero mean obeying 

\begin{align}
    J_{i_1\ldots i_{\frac{q}{2}}j_1\ldots j_{\frac{q}{2}}}^{(0, 1)} &= (J_{j_1\ldots j_{\frac{q}{2}}i_1\ldots i_{\frac{q}{2}}}^{(0, 1)})^*, \\
    \langle (J_{i_1\ldots i_{\frac{q}{2}}j_1\ldots j_{\frac{q}{2}}}^{(0, 1)})^2 \rangle &= \frac{J_{0, 1}^2 (q/2)!(q/2-1)!}{N^{q-1}}.
\end{align}
The numerical coefficients are chosen to cancel additional factors in the path integral and the powers of $N$ ensure the correct scaling of extensive thermodynamic variables. 

For future convenience, we write the total system Hamiltonian in terms of bond operators $H^{x, x+1}$ acting on consecutive sites $(x, x+1)$

\begin{equation}
    H = \sum_x H^{x, x+1} = \sum_x \left(\frac{1}{2}H_0^x+\frac{1}{2}H_0^{x+1}+H_1^{x, x+1}\right).
    \label{eq:ham}
\end{equation}
This Hamiltonian is invariant under the particle-hole symmetry  and has a globally conserved $U(1)$ charge density $Q=\sum_x Q_x$, where the local charge density $Q_x\in(-1/2, 1/2)$ is defined as in Ref.~\cite{davison2017}
\begin{equation}
    Q_x = \frac{1}{N}\sum_i \langle (c_i^x)^\dagger c_i^x \rangle - \frac{1}{2}.
    \label{eq:charge}
\end{equation}
The only other conserved quantity is the energy and it is precisely the interplay between the transport properties of these two conserved charges that we aim to study.  

We should mention that similar higher-dimensional SYK models have been previously studied in the context of transport~\cite{gu2017diffusion, gu2017,davison2017,cmjian2017,song2017,patel2018,chowdhury2018,guo2019,guo2020,zanoci2022}, quantum chaos~\cite{gu2017entanglement,chen2017,zhang2017,bentsen2019,chen2020}, and quantum phase transitions~\cite{banerjee2017,haldar2018,cmjian2017,jian2017mbl,cai2018}. Most importantly, a generalization of our setup to arbitrary graphs coupled to thermal reservoirs should be straightforward~\cite{zanoci2022}.

\section{Methods}
\label{sec:methods}

In this section, we derive the equations governing the equilibrium and non-equilibrium dynamics of our model, review the definitions of various thermodynamic, transport, and chaos-related quantities that we report later in the paper, and show how these observables simplify in the limit of small and large $q$. 

\subsection{Equilibrium}
\label{sec:equilibrium_methods}

We begin with the equilibrium description of our model, which is most easily done in imaginary time. The SYK chain maintains all the exactly solvable properties of a single SYK cluster in the large-$N$ limit~\cite{sachdev2015,davison2017,song2017,Gu2020}. We are interested in the grand canonical partition function $Z=\Tr e^{-\beta(H-\mu Q)}$, where $\beta=1/T$ is the inverse temperature and $\mu$ is the chemical potential that fixes the value of $Q$. In equilibrium, these parameters are constant (site-independent) throughout the system. Due to the self-averaging property of this model at large $N$~\cite{sachdev2015}, it is sufficient to consider the replica-diagonal partition function $Z=\int [\diff G_x][\diff \Sigma_x]e^{-S}$, for which the Euclidean effective action, after integrating out the fermions, becomes

\begin{gather}
    S = \sum_x (S_{x, x+1} + S_x), \\
    S_{x, x+1} = - \frac{2NJ_1^2}{q}\int \diff\tau_1 \diff\tau_2 (-G_x(\tau_1, \tau_2)
    G_{x+1}(\tau_2, \tau_1))^{\frac{q}{2}},
\end{gather}
\begin{widetext}
\begin{equation}
    S_{x} = -N\log\det \big((\partial_\tau-\mu)\delta(\tau_1-\tau_2) + \Sigma_x(\tau_1, \tau_2)\big)- N\int \diff\tau_1 \diff\tau_2\Big(\Sigma_x(\tau_1, \tau_2)G_x(\tau_2, \tau_1) + \frac{J_0^2}{q}(-G_x(\tau_1, \tau_2)G_x(\tau_2, \tau_1))^{\frac{q}{2}} \Big).
\end{equation}
\end{widetext}
For each site $x$, we defined the Euclidean time-ordered fermion two-point function
\begin{equation}
    G_x(\tau_1, \tau_2) = -\frac{1}{N} \sum_{i=1}^N \langle Tc_i^x(\tau_1)c_i^x(\tau_2)^\dagger \rangle,
\end{equation}
and the fermion self-energy $\Sigma_x(\tau_1, \tau_2)$ as the associated Lagrange multiplier. In the large-$N$ limit, the saddle point of this effective action produces the Schwinger-Dyson (SD) equations of motion 
\begin{align}
    G_x(i\omega_n) &= \frac{1}{i\omega_n+\mu-\Sigma_x(i\omega_n)},\\
    \Sigma_x(\tau) &=  (-1)^{\frac{q}{2}-1}G_x(\tau)^{\frac{q}{2}}\Big(J_0^2G_x(-\tau)^{\frac{q}{2}-1} \nonumber\\
    &+ J_1^2G_{x-1}(-\tau)^{\frac{q}{2}-1}+J_1^2G_{x+1}(-\tau)^{\frac{q}{2}-1}\Big), \label{eq:system_SD}
\end{align}
where $\omega_n=(2n+1)\pi/\beta$ is a Matsubara frequency and we assumed time-translation invariance $\tau=\tau_1-\tau_2$ in equilibrium. At half filling ($\mu=0$), the Green's function and all the other quantities derived from it are identical to those of a Majorana SYK model. The only difference is that complex fermions have twice as many degrees of freedom, leading to a trivial doubling of all the extensive observables. 

As we have previously shown~\cite{zanoci2022}, for a uniform chain in equilibrium, the Green's functions take on the site-independent value $G(\tau)$ and the effective on-site coupling become $J=\sqrt{J_0^2+2J_1^2}$. In other words, the SD equations for an interacting SYK cluster have the exact same form as those of an isolated $(0+1)$ dimensional SYK model with coupling $J$

\begin{equation}
\begin{split}
    G(i\omega_n) &= \frac{1}{i\omega_n+\mu-\Sigma(i\omega_n)},\\
    \Sigma(\tau) &=  J^2G(\tau)^{\frac{q}{2}}(-G(-\tau))^{\frac{q}{2}-1}. \label{eq:system_SD_2}
\end{split}
\end{equation}
This system of SD equations can be solved numerically using the method described in \appref{sec:appendixA}. 

\subsection{Thermodynamics}
\label{sec:thermo}

Once we obtain the solutions of the equilibrium SD equations, we can compute any thermodynamic variable. Our goal here is twofold. First, we would like to derive expressions for the common thermodynamic functions, such as entropy and heat capacity, that would help us identify a phase transition in the complex SYK model and assess its impact on the transport coefficients and chaos~\cite{azeyanagi2018,ferrari2019,Sorokhaibam2020,samui2021,tikhanovskaya2021a,cao2021thermodynamic}. Second, we need to find the susceptibility matrix, which relates diffusivities and conductivities~\cite{de1984non,kubo1985statistical,forster1975hydrodynamic,hartnoll2014}, as described in more detail in \secref{sec:transport_methods}.  

Since our system is homogeneous, we can focus on the thermodynamic properties of a single SYK node with interaction strength $J$. In what follows, all the extensive quantities are replaced by their densities per particle (i.e. divided by $N$). As is the case for most thermodynamic problems, our starting point is the grand canonical potential $\Omega$. In the large-$N$ limit, $\Omega$ is approximated by evaluating the action on the saddle-point equations of motion 

\begin{equation}
\begin{split}
    \Omega &= -T\log{Z}= T\Bigg[\sum_n \log\left(\frac{G(i\omega_n)}{G_0(i\omega_n)}\right) \\
    &-\frac{q-1}{q}\sum_n\Sigma(i\omega_n)G(i\omega_n)-\log\left(2\cosh(\mu/2T)\right)\Bigg],
    \label{eq:grand_canonical}
\end{split}
\end{equation}
where $G_0(i\omega_n) = (i\omega_n+\mu)^{-1}$ is the free fermion Green's function. Here we have regularized the logarithm by adding and subtracting the free fermion contribution~\cite{davison2017,song2017,Gu2020}, and evaluated the last term using the Matsubara frequency summation~\cite{de1984non,kubo1985statistical,forster1975hydrodynamic}. The free energy is given by a Legendre transform $F=\Omega+\mu Q$. Recall that the charge and chemical potential can be obtain from their respective ensembles at fixed temperature 

\begin{equation}
\begin{split}
    Q &= -\left(\frac{\partial\Omega}{\partial \mu}\right)_T = \frac{1}{2}\left(G(0^+)-G(\beta^-)\right),\\
    \mu &= \left(\frac{\partial F}{\partial Q}\right)_T = -\partial_\tau G(0^+)-\partial_\tau G(\beta^-),
    \label{eq:charge_equilibrium}
\end{split}
\end{equation}
where the second equalities in terms of Green's function are derived in Ref.~\cite{sachdev2015,Gu2020}. Note that our definition of the grand canonical potential in \eqref{eq:grand_canonical} gives us exactly the charge density introduced in \eqref{eq:charge}.

The entropy is computed as the first derivative of the potential, using the standard thermodynamic identities 

\begin{equation}
    S = -\left(\frac{\partial\Omega}{\partial T}\right)_\mu = -\left(\frac{\partial F}{\partial T}\right)_Q.
    \label{eq:entropy}
\end{equation}
A striking feature of the SYK model is its non-zero residual entropy $S_0$ in the limit of zero temperature, which is not due to an exponentially large ground state degeneracy, but rather because of the exponentially small level spacing all the way down to the ground state~\cite{parcollet1999,georges2000,georges2001,sachdev2015}. We will use this entropy to distinguish between an SYK-like ground state and a trivial one in \secref{sec:equilibrium_results}, and will also relate it to the thermopower in \appref{sec:appendixD}.

The static susceptibility matrix $\chi$ relates the change in macroscopic observables due to the variation of the underlying microscopic quantities

\begin{equation}
    \begin{pmatrix}
    \nabla Q\\
    \nabla E - \mu\nabla Q
    \end{pmatrix} = \begin{pmatrix}
    \chi_{11} & \chi_{12} \\
    \chi_{21} & \chi_{22}
    \end{pmatrix}
    \begin{pmatrix}
    \nabla \mu\\
    \nabla T
    \end{pmatrix},
    \label{eq:susceptibility}
\end{equation}
and can be cast in terms of the second derivative of the grand potential~\cite{de1984non,kubo1985statistical,forster1975hydrodynamic} 

\begin{equation}
    \chi = \begin{pmatrix}
    -\left(\frac{\partial^2\Omega}{\partial \mu^2}\right)_T & -\left(\frac{\partial^2\Omega}{\partial \mu \partial T}\right)_{\mu, T} \\
    -T\left(\frac{\partial^2\Omega}{\partial T \partial \mu}\right)_{T, \mu} & -T\left(\frac{\partial^2\Omega}{\partial T^2}\right)_\mu
    \end{pmatrix}.
    \label{eq:susceptibility_2nd_deriv}
\end{equation}
By virtue of equality of mixed partial derivatives, the off-diagonal elements are always related by $\chi_{21} = T\chi_{12}$. The diagonal elements can be interpreted as the charge compressibility 

\begin{equation}
    K \equiv \chi_{11} = -\left(\frac{\partial^2\Omega}{\partial \mu^2}\right)_T = \left(\frac{\partial Q}{\partial \mu}\right)_T,
    \label{eq:charge_compressibility}
\end{equation}
and heat capacity at constant chemical potential 

\begin{equation}
    C_\mu \equiv \chi_{22} = -T\left(\frac{\partial^2\Omega}{\partial T^2}\right)_\mu = T\left(\frac{\partial S}{\partial T}\right)_\mu.
    \label{eq:C_mu}
\end{equation}
The heat capacity at fixed charge can be related to the other entries in the susceptibility matrix via the thermodynamic identity~\cite{hartnoll2014}

\begin{equation}
    C_Q = T\left(\frac{\partial S}{\partial T}\right)_Q = C_\mu - \frac{T\chi_{12}^2}{K}.
\end{equation}
Finally, the linear-in-$T$ coefficient of the specific heat is simply defined as $\gamma = C_Q/T$. 

Both $K$ and $\gamma$ play an important role in transport. They appear as the coefficients in the low-temperature Schwarzian effective action used to describe charge and energy fluctuations~\cite{davison2017}. We will also show that in this conformal limit, the ratio of energy to charge diffusivities is governed by $K/\gamma$.

\subsection{Non-equilibrium}
\label{sec:nonequilibrium_methods}

Although the Euclidean time formulation works well for thermodynamics, it is not suitable for non-equilibrium dynamics, due to the problems arising from analytic continuation to zero frequency. Therefore, the non-equilibrium evolution of a quantum many-body system is better described in real-time using the Schwinger-Keldysh formalism~\cite{kamenev2011field,stefanucci2013nonequilibrium}. Following the derivation in Refs.~\cite{song2017,haldar2020}, we can write down the terms in a Lorentzian effective action after integrating out the fermions, just as we did in imaginary time
\begin{equation}
    S_{x, x+1}  = \frac{2iNJ_1^2}{q}\int_{\mathcal{C}} \diff t_1 \diff t_2 G_x(t_2, t_1)^{\frac{q}{2}}
    G_{x+1}(t_1, t_2)^{\frac{q}{2}},
\end{equation}
\begin{widetext}
\begin{equation}
    S_x  = -iN\log\det \left(\partial_t\delta_{\mathcal{C}}(t_1, t_2) + i\Sigma_x(t_1, t_2)\right)+ iN\int_{\mathcal{C}} \diff t_1 \diff t_2\left( \frac{J_0^2}{q}G_x(t_2, t_1)^{\frac{q}{2}}G_x(t_1, t_2)^{\frac{q}{2}}-\Sigma_x(t_1, t_2)G_x(t_2, t_1) \right),
\end{equation}
\end{widetext}
where $\mathcal{C}$ denotes the closed-time Keldysh contour consisting of a positive and a negative branch~\cite{kamenev2011field,stefanucci2013nonequilibrium}. Recall that in the Schwinger-Keldysh formalism, the contour-ordered Green's functions $G_x(t_1, t_2)$ are actually $2\times2$ matrices, where each entry corresponds to a placement of the time arguments on either branch of the contour. The off-diagonal entries of this matrix are the well-known greater and lesser Green's functions

\begin{align}
    G_x^>(t_1, t_2) \equiv G_x(t_1^-, t_2^+) &= -\frac{i}{N} \sum_{i=1}^N \langle c_i(t_1^-)c_i^\dagger(t_2^+) \rangle,\\
    G_x^<(t_1, t_2) \equiv G_x(t_1^+, t_2^-) &= \frac{i}{N} \sum_{i=1}^N \langle c_i^\dagger(t_2^-)c_i(t_1^+) \rangle,
\end{align}
where $t_i^\pm$ live on the positive and negative branch respectively. The contour-ordered Green's functions above are related to the more conventional retarded, advanced, and Keldysh Green's functions via a Keldysh rotation~\cite{kamenev2011field,stefanucci2013nonequilibrium}

\begin{align}
    G_x^R(t_1, t_2) &= \Theta(t_1-t_2)\big( G_x^>(t_1, t_2)-G_x^<(t_1, t_2)\big),\\
    G_x^A(t_1, t_2) &= \Theta(t_2-t_1)\big( G_x^<(t_1, t_2)-G_x^>(t_1, t_2)\big),\\
    G_x^K(t_1, t_2) &= G_x^>(t_1, t_2)+G_x^<(t_1, t_2),
\end{align}
The corresponding self-energies are defined in a similar manner.

For the purposes of our analysis, we will only consider states in thermal equilibrium or steady states weakly perturbed from equilibrium~\cite{zanoci2022}. In both cases, the fermion Green's functions become time-translation invariant and satisfy the identity~\cite{babadi2015} 

\begin{equation}
    G_x^\gtrless(t_1, t_2) = G_x^\gtrless(t=t_1-t_2) = -G_x^\gtrless(-t)^*.
    \label{eq:gtr_lsr}
\end{equation}
Furthermore, their values at $t=0$ are related to the local charge and chemical potential via 
\begin{equation}
\begin{split}
    Q_x &= -iG_x^>(0)+\frac{1}{2} = -iG_x^<(0)-\frac{1}{2}, \\
    \mu_x &= \partial_tG_x^>(0) - \partial_tG_x^<(0),
\end{split}
\label{eq:charge_real}
\end{equation}
which follow from an analytic continuation of \eqref{eq:charge_equilibrium} to real time. 

To obtain the Schwinger-Dyson equations governing the real-time dynamics of the system, we can look for large-$N$ saddle point solutions of the Lorentzian action~\cite{song2017,haldar2020}

\begin{equation}
\begin{split}
    G_x^R(\omega) &= \frac{1}{i\omega - \Sigma_x^R(\omega)},\\
    \Sigma_x^\gtrless(t) &= G_x^\gtrless(t)^{\frac{q}{2}}\Big(J_0^2 G_x^\lessgtr(-t)^{\frac{q}{2}-1} \\
    &+ J_1^2G_{x-1}^\lessgtr(-t)^{\frac{q}{2}-1} + J_1^2G_{x+1}^\lessgtr(-t)^{\frac{q}{2}-1}\Big).
     \label{eq:SD_real_time}
\end{split}
\end{equation}
Note that these equations are only valid for the time-translation invariant case. For more general non-equilibrium setups, one has to derive the full Kadanoff-Baym equations and solve them numerically~\cite{zanoci2022}. 

We emphasize that the real-time action does not involve a chemical potential, since $\mu_x$ is a property of the state, rather than the Hamiltonian. Often times this issue is addressed by explicitly adding a mass term $-\mu_x\sum_{i=1}^N (c_i^x)^\dagger c_i^x$ to the Hamiltonian~\cite{sachdev2015,song2017,can2019,haldar2020,cheipesh2020}. This works in imaginary time, where it is simply equivalent to working in the grand canonical ensemble. However, the chemical potential and mass term are not equivalent in real time~\cite{Sorokhaibam2020}, with the Green's functions being typically off by a factor of $e^{i\mu_x t}$, which can lead to incorrect dynamics and transport properties. 

Similarly, \eqref{eq:SD_real_time} does not have an explicit dependence on chemical potential or temperature, in contrast to its Euclidean-time counterpart in \eqref{eq:system_SD}. Hence there are infinitely many distinct saddle point solutions of the real-time SD equation to which we can converge, each corresponding to different $\mu_x$ and $\beta_x$ . To circumvent this problem, we use the fluctuation-dissipation theorem (FDT)~\cite{de1984non,kubo1985statistical,forster1975hydrodynamic} to set the values of these parameters

\begin{equation}
    \frac{iG_x^K(\omega)}{A_x(\omega)} = \tanh(\frac{\beta_x(\omega-\mu_x)}{2}),
    \label{eq:fdt}
\end{equation}
where $A_x(\omega) = -2\Im G_x^R(\omega)$ is the spectral function and $G_x^K(\omega)$ is the Fourier transform of the Keldysh Green's function. Fixing the local temperature and chemical potential in this way is applicable to both the equilibrium and near-thermal steady states under consideration~\cite{zanoci2022}. The FDT, together with \eqref{eq:SD_real_time}, form a closed set of equations that can be solved iteratively to find a unique solution (see \appref{sec:appendixA}).

So far we have assumed that each cluster has a well defined chemical potential $\mu_x$ and inverse temperature $\beta_x$. This is indeed true for a uniform chain in equilibrium with $\mu_x=\mu$ and $\beta_x=\beta$. Similarly to the imaginary-time version, the real-time site-independent solution $G^\gtrless(t)$ is the same as that of a single SYK node with an effective coupling $J=\sqrt{J_0^2+2J_1^2}$. We have also shown that in the presence of a small bias throughout the chain, the NESS Green's functions at late times are only slightly perturbed from their equilibrium values~\cite{zanoci2022}, as long as we are still in the linear response regime. This bias can be introduced by either directly coupling the system to baths at different chemical potentials and temperatures~\cite{chen2017,zhang2019,almheiri2019,cheipesh2020,haldar2020,zanoci2022,can2019}, or by introducing an effective coupling to the environment through Lindblad operators~\cite{sa2021lindbladian,kulkarni2021syk}. As we will discuss in detail in the next section, to extract the thermoelectric transport coefficients, it is enough to impose a uniform chemical potential or temperature gradient along the chain. Thus one can define local parameters that are ever so slightly perturbed from their equilibrium values

\begin{equation}
\begin{split}
    \mu_x &= \mu + x\nabla\mu, \\
    \beta_x &= \beta + x\nabla\beta,
\end{split}
\label{eq:beta_mu_grad}
\end{equation}
with $|\nabla\mu|\ll\mu$ and $|\nabla\beta|\ll\beta$. This allows us to write the near-equilibrium Green's functions in terms of an extra site-dependent correction 

\begin{equation}
    G_x^\gtrless(t) = G^\gtrless(t) + xF_{\mu,\beta}^\gtrless(t),
    \label{eq:ness_ansatz}
\end{equation}
where $|F_{\mu, \beta}^\gtrless(t)|\ll |G^\gtrless(t)|$ are the non-equilibrium contributions proportional to the gradients

\begin{equation}
\begin{split}
    F_\mu^\gtrless(t) &= \frac{\diff G^\gtrless(t)}{\diff \mu}\nabla\mu, \\
    F_\beta^\gtrless(t) &= \frac{\diff G^\gtrless(t)}{\diff \beta}\nabla\beta.
    \label{eq:def_F}
\end{split}
\end{equation}
The subscripts refer to whether the perturbation is due a chemical potential or a temperature gradient. To first order, these contributions can be summed to characterize the system's response to any mixed thermoelectric bias. \eqref{eq:ness_ansatz} is analogous to a gradient expansion in hydrodynamics. We see that to access the non-equilibrium transport physics in the linear response regime, it is sufficient to solve the SD equations in real time at equilibrium. 

\begin{figure}[htp]
	\begin{center}
	\includegraphics[width = \columnwidth]{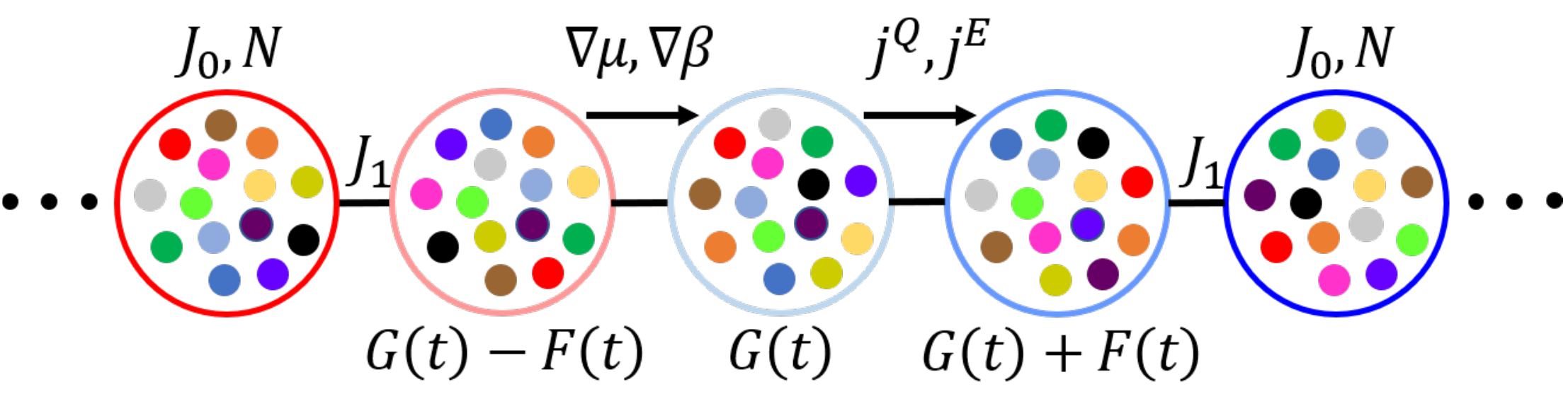}
	\caption{Schematic depiction of the infinite one-dimensional SYK chain in non-equilibrium. Each site contains $N$ fermions with intra-cluster coupling $J_0$ and inter-cluster coupling $J_1$. The system is subject to uniform biases $\nabla\mu$ and $\nabla\beta$. In the steady state, homogeneous charge and energy currents $j^{Q,E}$ flow through the chain. The local Green's functions are perturbed from their equilibrium values $G(t)$ by the linear response contributions $F(t)$.}
	\label{fig:fig1}
	\end{center}
\end{figure}

\subsection{Transport}
\label{sec:transport_methods}

Our model has charge and energy as the only two conserved quantities. These are expressed in terms of the local on-site charge density $Q_x$ and the on-bond energy density $E_{x, x+1}=\langle H^{x, x+1}\rangle = (E_0^x+E_0^{x+1})/2 + E_1^{x, x+1}$ introduced in \secref{sec:model}. The charge density can be computed from the real-time Green's functions using \eqref{eq:charge_real}, while the energy density has contributions from 
\begin{align}
    E_0^x &=  -2i\frac{J_0^2}{q}\int_{-\infty}^{t}\diff t_1 \left(G_x^>(t, t_1)^{\frac{q}{2}} G_x^<(t_1, t)^{\frac{q}{2}} - \hc \right), \nonumber\\
    E_1^{x, x+1} &= -2i\frac{J_1^2}{q}\int_{-\infty}^{t}\diff t_1 \Big(G_x^>(t, t_1)^{\frac{q}{2}} G_{x+1}^<(t_1, t)^{\frac{q}{2}} \nonumber\\
    &+ G_x^<(t_1, t)^{\frac{q}{2}} G_{x+1}^>(t, t_1)^{\frac{q}{2}} - \hc \Big).
\end{align}
These conserved quantities have an associated charge current density $j_x^Q = \diff Q_x/\diff t$ and energy current density $j_x^E = \diff E_{x, x+1}/\diff t$ respectively. The formulas for the currents flowing across a site $x$ can be derived by combining the continuity equation and Heisenberg's equation of motion~\cite{zotos1997transport,kapustin2021}, resulting in 
\begin{align}
    j_x^Q &= i[Q_x, H^{x, x+1}],\\
    j_x^E &= i[H^{x-1, x}, H^{x, x+1}].
\end{align}
Computing the expectation value of these commutators in the Schwinger-Keldysh formalism is more involved and we provide a derivation in \appref{sec:appendixB}. Our general formulas for the currents are given by Eqs.~(\ref{eq:j_Q},~\ref{eq:j_E}-\ref{eq:j_+-}).

In equilibrium, there are no currents flowing through the system. To observe a finite current, we have to introduce a small bias, accomplished, for instance, by connecting the chain to reservoirs at its two ends~\cite{zanoci2022}. In the long-time limit, when the system reaches its steady state, the currents become uniform throughout the chain $j^{Q, E}\equiv\langle j_x^{Q, E}\rangle$, as shown in \figref{fig:fig1}. In the linear response regime, the gradients are also small and constant as in \eqref{eq:beta_mu_grad}. Therefore, we can use \eqref{eq:ness_ansatz} to write all the quantities of interest in terms of the equilibrium Green's functions and to first order in non-equilibrium corrections $F_{\mu, \beta}^\gtrless(t)$. For example, the charge gradient becomes 
\begin{equation}
    \nabla Q = -iF_{\mu,\beta}^>(0),
    \label{eq:grad_Q}
\end{equation}
while the energy gradient is given by 

\begin{widetext}
\begin{equation}
    \nabla E = 2J^2 \int_0^{\infty} \diff t \im\left[\left(-G^>(t)G^<(t)^*\right)^{\frac{q}{2}}\left(\frac{F_{\mu,\beta}^>(t)}{G^>(t)}+\frac{F_{\mu,\beta}^<(t)^*}{G^<(t)^*}\right)\right].
    \label{eq:grad_E}
\end{equation}
Similarly, the charge current in \eqref{eq:j_Q} becomes

\begin{equation}
     j^Q = \frac{qJ^2}{2} \int_0^{\infty} \diff t \re\left[\left(-G^>(t)G^<(t)^*\right)^{\frac{q}{2}}\left(\frac{F_{\mu,\beta}^>(t)}{G^>(t)}-\frac{F_{\mu,\beta}^<(t)^*}{G^<(t)^*}\right)\right],
    \label{eq:j_Q_linear}
\end{equation}
and the energy current in Eqs.~(\ref{eq:j_E}-\ref{eq:j_+-}) simplifies to 

\begin{equation}
    j^E = \frac{1}{2}J_1^2J^2\Re(j_{++}+j_{+-}),
\end{equation}
\begin{align}
    j_{++} &= q \int_{0}^{\infty}\diff t \int_{t}^{\infty} \diff t' \left(G^>(t)G^<(t)^*G^>(t')G^<(t')^*\right)^{\frac{q}{2}-1}\nonumber \\
    &\cdot \Big(G^<(t'-t)^*\left(G^<(t)^*F_{\mu,\beta}^>(t') - G^>(t')F_{\mu,\beta}^<(t)^*\right) - G^>(t'-t)\left(G^<(t')^*F_{\mu,\beta}^>(t) - G^>(t)F_{\mu,\beta}^<(t')^*\right)\Big),\\
    j_{+-} &= -q \int_{0}^{\infty}\diff t \int_{t}^{\infty} \diff t' \left(G^>(t)^*G^<(t)G^>(t')G^<(t')^*\right)^{\frac{q}{2}-1}\nonumber \\
    &\cdot \Big(G^<(t'-t)^*\left(G^>(t)^*F_{\mu,\beta}^>(t') - G^>(t')F_{\mu,\beta}^>(t)^*\right) - G^>(t'-t)\left(G^<(t')^*F_{\mu,\beta}^<(t) - G^<(t)F_{\mu,\beta}^<(t')^*\right)\Big).
\end{align}
\end{widetext}

Within linear response, the currents are related to the conjugate gradients via the conductivity matrix $L$

\begin{equation}
    \begin{pmatrix}
    j^Q\\
    j^E - \mu j^Q
    \end{pmatrix} = - \begin{pmatrix}
    \sigma & \alpha \\
    \alpha T & \bar{\kappa}
    \end{pmatrix}
    \begin{pmatrix}
    \nabla \mu\\
    \nabla T
    \end{pmatrix},
    \label{eq:conductivity}
\end{equation}
where $\sigma\equiv L_{11}$ is the electrical conductivity, $\alpha\equiv L_{12}$ is the thermoelectric conductivity, and $\kappa \equiv L_{22} - L_{12}L_{21}/L_{11} = \bar{\kappa} - \alpha^2 T/\sigma$ is the thermal conductivity~\cite{de1984non,kubo1985statistical,forster1975hydrodynamic,hartnoll2014}. The off-diagonal elements are constrained by the Onsager reciprocal relation $L_{21} = TL_{12}$. The quantity $j^E - \mu j^Q$ is referred to as the heat current~\cite{hartnoll2014}. \eqref{eq:conductivity} contains three unknown transport coefficients. To solve it, we will consider two different setups (see \figref{fig:fig1}): one with $\nabla \mu = \text{const}$ and $\nabla T = 0$ (or equivalently $\nabla \beta = 0$), and the other with $\nabla \mu = 0$ and $\nabla T = \text{const}$. This will give us a system of equations from which we can easily derive $\sigma$, $\alpha$, and $\bar{\kappa}$ (or $\kappa$). We will refer to the non-equilibrium contribution in each scenario as $F_\mu^\gtrless(t)$ and $F_\beta^\gtrless(t) $ respectively (see \eqref{eq:def_F}). 

The SYK model is known to exhibit diffusive transport~\cite{davison2017}. The hydrodynamic relations defining the diffusivity matrix $D$ are given by 

\begin{equation}
    \begin{pmatrix}
    j^Q\\
    j^E - \mu j^Q
    \end{pmatrix} = - \begin{pmatrix}
    D_{11} & D_{12} \\
    D_{21} & D_{22}
    \end{pmatrix}
    \begin{pmatrix}
    \nabla Q\\
    \nabla E - \mu\nabla Q
    \end{pmatrix},
    \label{eq:diffusivity}
\end{equation}
where $D_{11}$ is the charge diffusion constant, $D_{22}$ is the thermal (not energy!) diffusion constant, and the off-diagonal elements describe mixed transport~\cite{hartnoll2014}. The diffusivity matrix can be diagonalized, with eigenvalues $D_\pm$ describing the coupled diffusion of charge and heat. It is these modes that govern the dynamics of charge and energy fluctuations in the system and are thus more physically relevant than the individual entries in $D$~\cite{hartnoll2014,davison2017}. In particular, in order for the fluctuations to decay, we must have that $D_{\pm}\geq 0$, while the matrix elements $D_{ij}$ can be negative.  

Finally, combining Eqs.~(\ref{eq:susceptibility},~\ref{eq:conductivity},~\ref{eq:diffusivity}) yields the generalized Einstein relation $L=D\chi$. In the absence of coupling between the charge and energy carriers (e.g. at $\mu = 0$), we have $\chi_{12} = \alpha = 0$ and recover the standard Einstein relations for charge $\sigma = D_{11}K$ and energy $\kappa = D_{22}C_\mu$ transport~\cite{hartnoll2014}. However, more generally, one has the coupled relations given by the full matrix equation. 

\subsection{Chaos}
\label{sec:chaos_methods}

The non-Fermi liquid phase described by the SYK model is known to be highly chaotic~\cite{kitaev2015simple,maldacena2016} and even saturates a bound on chaos at low temperatures~\cite{maldacena2016chaos}. In such maximally chaotic theories, energy dynamics and diffusion are fundamentally related to chaos~\cite{Blake2018,Blake2021}. Moreover, the thermal diffusion constant of SYK models in the conformal limit is directly controlled by the butterfly velocity~\cite{gu2017diffusion,davison2017,zanoci2022,choi2021}, thus realizing a conjectured bound on diffusion in incoherent metals~\cite{hartnoll2014,blake2016_1,blake2016_2,blake2017,hartman2017,gu2017,chen2020,choi2021,hartnoll2021planckian,Blake2018,Blake2021}. In this section, we review the many-body chaos properties of the SYK model and will later show that chaos provides an upper bound on diffusivity in SYK chains at any temperature and chemical potential. We will mostly follow our analysis of the Majorana SYK model~\cite{zanoci2022}.

We begin by introducing the out-of-time-order correlation function (OTOC), which has been widely used as a measure of chaos in quantum systems~\cite{larkin1969,shenker2014,shenker2015,maldacena2016chaos,maldacena2016,bryce2021}. The regularized OTOC in real time is defined as

\begin{equation}
    C(x, t_1, t_2) = \frac{1}{N^2}\sum_{i, j=1}^N \Tr[yc_j^x(t_1)^\dagger yc_i^0(0)^\dagger yc_j^x(t_2)yc_i^0(0)],
\end{equation}
where $y=e^{-\beta H/4}/Z^{1/4}$ evenly spaces the fermionic fields along the thermal circle~\cite{guo2019}. To leading order, the OTOC can be written as  

\begin{equation}
    C(x, t_1, t_2) = \mathcal{F}_d-\frac{\mathcal{F}(x, t_1, t_2)}{N},
\end{equation}
where $\mathcal{F}_d$ is just a constant corresponding to the disconnected correlator and $\mathcal{F}$ is the first order contribution stemming from the contraction of $c_i^0$ with $c_j^x$~\cite{mezei2020}. For a chaotic system with a large
hierarchy of timescales between thermalization and scrambling, we expect $\mathcal{F}$ to scale exponentially as $e^{\lambda_Lt}$, where $t_1=t_2=t$ is in the Lyapunov regime $\beta\lesssim t\lesssim \beta\ln N$~\cite{mezei2020}. The Lyapunov exponent $\lambda_L$ determines the rate of growth of an operator under Heisenberg evolution and serves as a quantum mechanical measure for information scrambling in phase space~\cite{shenker2015}. For an isolated SYK cluster, $\mathcal{F}(t_1, t_2)$ is determined by summing over a set of ladder diagrams~\cite{maldacena2016,banerjee2017,zhang2017,Bulycheva2017,Bhattacharya2017,guo2019,Sorokhaibam2020}, resulting in the self-consistency equation 

\begin{equation}
    \mathcal{F}(t_1, t_2) = \int_{-\infty}^\infty \diff t_3 \diff t_4 K^R(t_1, t_2, t_3, t_4) \mathcal{F}(t_3, t_4),
\end{equation}
where $K_R$ is the retarded kernel 

\begin{widetext}
\begin{equation}
\begin{split}
    K^R(t_1, t_2, t_3, t_4) &= (q-1)J^2G^R(t_1-t_3)G^A(t_4-t_2)G^W(t_3-t_4)^{\frac{q}{2}-1}G^W(t_4-t_3)^{\frac{q}{2}-1}\\
    &= (q-1)J^2G^R(t_1-t_3)G^R(t_2-t_4)^*|G^W(t_3-t_4)|^{q-2}.
    \label{eq:kernel}
\end{split}
\end{equation}
\end{widetext}
Here $G^W$ is the Wightman Green's function and we used the symmetry properties $G^A(-t)=G^R(t)^*$ and $G^W(-t)=G^W(t)^*$ to simplify the expression. For fermionic systems, the Wightman propagator is related to the spectral function in frequency space via~\cite{guo2019}

\begin{equation}
    G^W(\omega) = \frac{A(\omega)}{2\cosh(\beta\omega/2)}.
    \label{eq:wightman}
\end{equation}

To determine the Lyapunov exponent, we follow the prescription in Ref.~\cite{gu2019}, which works for both Majorana and complex SYK. We define a variant of the kernel with a parameter $\alpha < 0$

\begin{equation}
    K_\alpha^R(t, t') = \int_{-\infty}^\infty \diff s e^{\alpha s} K^R\left(s+\frac{t}{2}, s-\frac{t}{2}, \frac{t'}{2}, -\frac{t'}{2}\right).
    \label{eq:kernel_alpha}
\end{equation}
This operator can be cast in matrix form, with its largest eigenvalue $k_R(\alpha)$ depending on $\alpha$. The Lyapunov exponent is then determined by the equation $k_R(-\lambda_L)=1$. This condition is equivalent to $\mathcal{F}$ being an eigenvector of the kernel $K^R$ with eigenvalue one. The Lyapunov exponent of a $(0+1)$-d SYK model is known to saturate the bound $\lambda_L \leq 2\pi/\beta$ at low temperatures~\cite{kitaev2015simple,maldacena2016,maldacena2016chaos}. 

For spatially extended systems, such as our one-dimensional chain, the operators can also grow in space. Chaos propagation in a translation-invariant system is described by the Fourier transform of the momentum-space OTOC

\begin{equation}
    \mathcal{F}(x, t) \sim \int_{-\infty}^\infty \frac{\diff p}{2\pi} \frac{e^{\lambda_L(p)t+ipx}}{\cos(\lambda_L(p)\beta/4)},
\end{equation}
where $\lambda_L(p)$ is the momentum-dependent Lyapunov exponent~\cite{guo2019}. In the hydrodynamic limit, this integral can be evaluated using a saddle point approximation. Depending on the parameters of our model, the integral can either pick up a contribution solely from the saddle point $p_s$, or from both the saddle point and the momentum-space pole $p_1$, both of which are located on the imaginary axis $p_{s, 1}=i|p_{s, 1}|$~\cite{gu2019,guo2019}. In either case, the result can be written as

\begin{equation}
    \mathcal{F}(x, t) \sim e^{\lambda_L(p_{s, 1})(t-|x|/v_B)},
\end{equation}
where the butterfly velocity $v_B$ is defined as 

\begin{equation}
    v_B = \begin{cases} \frac{\lambda_L(p_s)}{|p_s|} &\mbox{if } |p_s|<|p_1| \\
            \frac{2\pi}{\beta |p_1|} & \mbox{if } |p_s| > |p_1| \end{cases}
\end{equation}
Physically, the butterfly velocity $v_B$ defines a light-cone that bounds the speed of operator growth in space~\cite{shenker2014}. It can also be viewed as a temperature-dependent extension of the Lieb-Robinson velocity~\cite{roberts2016}.

To compute the butterfly velocity, it is enough to find the momenta $p_{s, 1}$ according to Ref.~\cite{gu2019}. For a uniform SYK chain, the retarded kernel in momentum space factorizes $K^R(p) = s(p)K^R$, where $s(p)=1+\frac{qJ_1^2}{2(q-1)J^2}(\cos(p)-1)$ is the spatial kernel~\cite{gu2017diffusion} and $K^R$ is the kernel for a single cluster defined in \eqref{eq:kernel}. Therefore the eigenvalues of the kernel also factorize $k_R(p, \alpha) = s(p)k_R(\alpha)$. The momentum-dependent Lyapunov exponent can be obtained by solving the equation $k_R(p, -\lambda_L(p))=1$. At $p=0$, we recover our previous formula for the Lyapunov exponent of a single cluster. Generally, this equation has to be solved numerically by repeatedly diagonalizing the kernel in \eqref{eq:kernel_alpha} and using the bisection method, although closed-form solutions are available in some limits (see \secref{sec:large_q_methods}). Once we have the entire function $\lambda_L(p)$, the location of the saddle can then be found by solving $\lambda_L(p_s)=p_s\lambda_L^{'}(p_s)$, while $p_1$ is the momentum at which the Lyapunov exponent attains its maximum value $\lambda_L(p_1)=2\pi/\beta$. 

We can now use the newly introduced measures of chaos to define a characteristic chaos diffusivity $v_B^2/\lambda_L$, which is known to be closely related to the thermal diffusion constant in strange metals~\cite{davison2017,gu2017diffusion,zanoci2022,hartnoll2014,blake2016_1,blake2016_2,blake2017,hartman2017,gu2017,chen2020,choi2021,hartnoll2021planckian,Blake2018,Blake2021}. In fact, we will show that for all systems under consideration, the chaos diffusivity provides an upper bound $D_\pm\leq v_B^2/\lambda_L$, just as in the Majorana case~\cite{zanoci2022}.

\subsection{\texorpdfstring{$q=2$}{q=2} limit}
\label{sec:q=2_methods}

Although the Green's functions generally do not have a closed-form representation, they simplify significantly in the limits of small and large $q$, which we discuss next. We start with the special case of $q=2$ corresponding to free fermions, where the system has a quasiparticle description~\cite{davison2017} and the Hamiltonian becomes integrable and non-chaotic~\cite{garcia2018,haque2019}. 

The SD equations are quadratic and can be solved exactly~\cite{eberlein2017}. The spectral function in equilibrium is given by

\begin{equation}
    A(\omega) = \frac{2}{J}\sqrt{1-\left(\frac{\omega}{2J}\right)^2} \quad \mathrm{for} \ |\omega|<2J,
    \label{eq:spectral}
\end{equation}
and the Green's functions can be obtained from the fluctuation-dissipation theorem 

\begin{equation}
    G^\gtrless(\omega) = \mp \frac{iA(\omega)}{1+e^{\mp\beta(\omega-\mu)}},
    \label{eq:fdt_2}
\end{equation}
followed by an inverse Fourier transform 

\begin{equation}
    G^\gtrless(t) = \mp \frac{i}{\pi J}\int_{-2J}^{2J}\diff\omega \frac{e^{-i\omega t}}{1+e^{\mp\beta(\omega-\mu)}}\sqrt{1-\left(\frac{\omega}{2J}\right)^2}.
    \label{eq:green_q=2}
\end{equation}
Finally, we can derive the non-equilibrium contributions from \eqref{eq:def_F}. It is straightforward to check that $F_{\mu,\beta}^<(t) = F_{\mu,\beta}^>(t)$ and 

\begin{align}
    \Re\left[G^<(t) - G^>(t)\right] &= 0,\\
    \Im\left[G^<(t) - G^>(t)\right] &= \frac{B_1(2Jt)}{Jt},
\end{align}
where $B_1$ is the Bessel function of the first kind. With this in mind, we arrive at the following simplified formulas for the energy gradient and currents

\begin{align}
    &\nabla E = 2J\int_0^\infty \diff t \frac{B_1(2Jt)}{t}\Re F_{\mu,\beta}^>(t),\label{eq:grad_E_q=2}\\
    j^Q &= -\frac{J_1^2}{J}\int_0^\infty \diff t \frac{B_1(2Jt)}{t}\Im F_{\mu,\beta}^>(t),\label{eq:j_Q_q=2}\\
    j^E &= -J_1^2 \int_0^\infty \diff t \int_t^\infty \diff t' \frac{B_1(2Jt)}{t}\frac{B_1(2J(t'-t))}{t'-t}\Re F_{\mu,\beta}^>(t').
    \label{eq:j_E_q=2}
\end{align}
The charge gradient is still given by \eqref{eq:grad_Q}. The equations above contain all the necessary information to calculate the conductivities and diffusivities numerically at arbitrary $\mu$ and $\beta$ using the non-equilibrium setups described in \secref{sec:transport_methods}. Moreover, we were able to compute closed-form results for these transport coefficients in the limit of zero and infinite temperatures, as described in \appref{sec:appendixC}. 

\subsection{Large \texorpdfstring{$q$}{q} limit}
\label{sec:large_q_methods}

We now turn to the opposite limit of large $q$, where an analytic approximation for the Green's function at all temperatures is available~\cite{maldacena2016,davison2017,Bhattacharya2017,tarnopolsky2019}. To leading order in $1/q$, the Green's functions for an SYK model in equilibrium can be expanded as

\begin{align}
    G^>(t) &= -\frac{i}{e^{\beta\mu}+1}\left(1+\frac{g(t)}{q}+\cdots\right)\approx -\frac{ie^{g(t)/q}}{e^{\beta\mu}+1} , \\
    G^<(t) &= \frac{i}{e^{-\beta\mu}+1}\left(1+\frac{g(t)^*}{q}+\cdots\right)\approx \frac{ie^{g(t)^*/q}}{e^{-\beta\mu}+1},
\end{align}
where ``\ldots" denotes higher order terms, $g(t)$ is a function of order one satisfying $g(t)^* = g(-t)$ and $g(0) = 0$~\cite{Bhattacharya2017}. With this ansatz, the SD equations are equivalent to a differential equation for $g(t)$

\begin{equation}
    -\frac{\partial^2 g(t)}{\partial t^2} = iq\left(\Sigma^>(t)+\Sigma^<(t)^*\right) = 2\J^2e^{g(t)},
    \label{eq:g_diff_eq}
\end{equation}
where $\J = J\sqrt{q2^{1-q}\cosh^{2-q}(\beta\mu/2)}$ is the effective coupling. Notice that the original theory has two independent scales $\beta J$ and $\beta \mu$, while the new differential equation only depends on the combined $\beta\J$~\cite{Bhattacharya2017}. This holds even after including higher-order terms in $1/q$ and seems to be an artefact of this expansion~\cite{tarnopolsky2019}. The large $q$ limit is well defined only when we adjust the original coupling $J$ such that the re-scaled interaction $\J$ is kept finite as $q\to\infty$. This implies that $J$ has to be a function of $\beta\mu$, which makes the comparison with the numerical results at finite $q$ and constant $J$ a bit more complicated. A direct comparison is possible for $\mu=0$, where we recover some of our findings for Majorana fermions~\cite{zanoci2022}. 

The solution to \eqref{eq:g_diff_eq} is of the form found in Ref.~\cite{maldacena2016}

\begin{equation}
    e^{g(t)} = \frac{\cos^2(\pi v/2)}{\cosh^2\left(\frac{\pi v}{\beta} \left(\frac{i\beta}{2}+t\right)\right)},
    \label{eq:g}
\end{equation}
where $v$ satisfies

\begin{equation}
     \beta \J = \frac{\pi v}{\cos(\pi v/2)}.
     \label{eq:v}
\end{equation}
This gives us the full dependence of the Green's functions on $\mu$ and $\beta$. We find it convenient to write the derivatives $F_{\mu,\beta}^\gtrless$ as follows

\begin{align}
    \frac{F_{\mu,\beta}^>(t)}{G^>(t)} &= \frac{f_{\mu,\beta}(t)}{q}-\frac{C_{\mu,\beta}}{1+e^{-\beta\mu}},\\
    \frac{F_{\mu,\beta}^<(t)}{G^<(t)} &= \frac{f_{\mu,\beta}(t)^*}{q}+\frac{C_{\mu,\beta}}{1+e^{\beta\mu}},
\end{align}
where $C_\mu = \beta\nabla\mu$, $C_\beta=\mu\nabla\beta$, and $f_{\mu,\beta}(t)$ are non-equilibrium contributions to $g(t)$ in the presence of small gradients (see Ref.~\cite{zanoci2022} for the Majorana case)

\begin{equation}
\begin{split}
    f_\mu(t) &= \frac{\diff g(t)}{\diff \mu}\nabla\mu, \\
    f_\beta(t) &= \frac{\diff g(t)}{\diff \beta}\nabla\beta.
    \label{eq:def_f}
\end{split}
\end{equation}
Our non-equilibrium observables simplify drastically in terms of these functions 

\begin{widetext}
\begin{align}
    \nabla Q &= \frac{C_{\mu,\beta}}{4\cosh^2(\beta\mu/2)},\label{eq:grad_Q_large_q}\\
    \nabla E &= \frac{\J^2}{q^2\cosh^2(\beta\mu/2)} \int_0^{\infty} \diff t \im\left[e^{g(t)}\left(2f_{\mu,\beta}(t)-qC_{\mu,\beta}\tanh(\beta\mu/2)\right)\right],\label{eq:grad_E_large_q}\\
    j^Q &= -\frac{\J_1^2C_{\mu,\beta}}{4\cosh^2(\beta\mu/2)} \int_0^{\infty} \diff t \re\left[e^{g(t)}\right] = -\frac{\J_1^2C_{\mu,\beta}\cos(\pi v/2)}{4\J\cosh^2(\beta\mu/2)},\label{eq:j_Q_large_q}\\
    j^E &= \frac{1}{2}\J_1^2\J^2\Re(j_{++}+j_{+-}),\\
    j_{++} &= \frac{i}{q^2\cosh^2(\beta\mu/2)} \int_{0}^{\infty}\diff t \int_{t}^{\infty} \diff t' e^{g(t)+g(t')}\left(f_{\mu,\beta}(t')-f_{\mu,\beta}(t)-qC_{\mu,\beta}\tanh(\beta\mu/2)\right),\label{eq:j_++_large_q}\\
    j_{+-} &= \frac{i}{q^2\cosh^2(\beta\mu/2)} \int_{0}^{\infty}\diff t \int_{t}^{\infty} \diff t' e^{g(t)^*+g(t')}\left(f_{\mu,\beta}(t')-f_{\mu,\beta}(t)^*\right).
    \label{eq:j_+-_large_q}
\end{align}
\end{widetext}
Notice that the charge gradient and current do not have any dependence on $f_{\mu,\beta}(t)$ and we can already obtain closed-form expressions for them. On the other hand, the energy gradient and current require an explicit calculation of $f_{\mu,\beta}(t)$ for different biases, which we defer to \appref{sec:appendixC}. Nevertheless, we managed to compute the diffusivity and conductivity matrices analytically for arbitrary $\mu$ and $\beta$ in the large $q$ limit and will present our results in the next section. 

Finally, we comment on the chaos characteristics in this approximation. The Lyapunov exponent has been previously computed in the large $q$ limit~\cite{Bhattacharya2017,gu2019}. Since the same derivation applies for both Majorana and complex fermions, one finds, in our notation 

\begin{equation}
    k_R(\alpha) = \frac{8(\pi v)^2}{\alpha(\alpha-2\pi v/\beta)\beta^2}.
\end{equation}
Hence the momentum-dependent Lyapunov exponent is given by 

\begin{equation}
\begin{split}
    \lambda_L(p) &= \frac{\pi v}{\beta}\left(\sqrt{1+8s(p)}-1\right) \\
    &= \frac{\pi v}{\beta}\left(\sqrt{9+4\frac{J_1^2}{J^2}\left(\cos(p)-1\right)}-1\right),
\end{split}
\end{equation}
from which the butterfly velocity can be found numerically by solving the appropriate equations from \secref{sec:chaos_methods}. For a single SYK cluster we recover the well-known answer $\lambda_L(0) = 2\pi v/\beta$~\cite{Bhattacharya2017,gu2019}. We immediately see that in the low-temperature limit $v\to1$ and the system is maximally chaotic~\cite{kitaev2015simple,maldacena2016,maldacena2016chaos}. Moreover, in this limit, the butterfly velocity approaches $v_B = \pi\J_1/\sqrt{3\beta\J}$ and the thermal diffusion constant saturates the chaos bound $D_+=D_{22}=v_B^2/\lambda_L = \pi\J_1^2/6\J$ (see \eqref{eq:D_+_large_q}).  

\section{Results}
\label{sec:results}

\begin{figure}[tp]
	\begin{center}
	\includegraphics[width = \columnwidth]{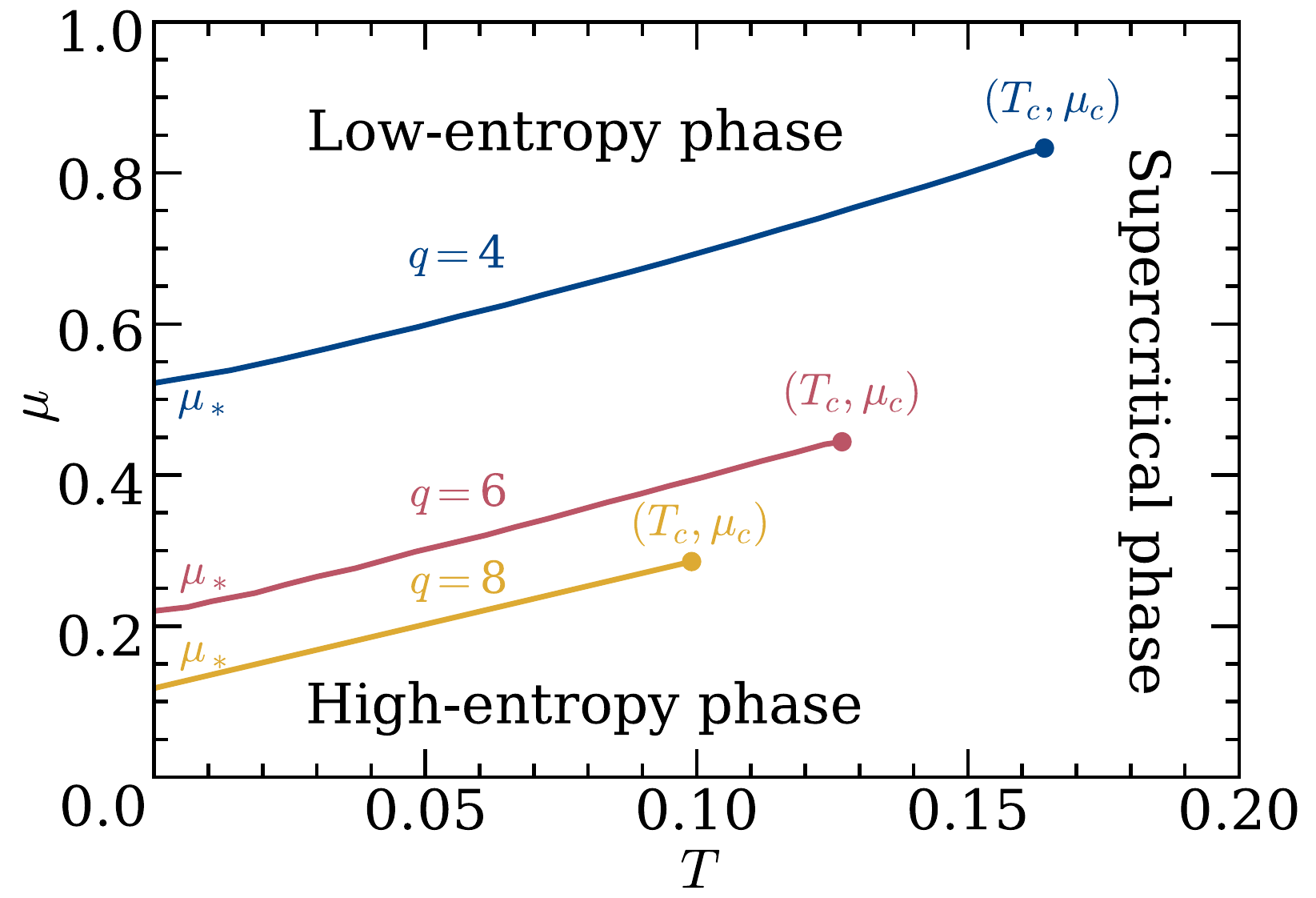}
	\caption{Phase diagram of the complex SYK model for $q=4-8$, in units of $J=\sqrt{3\cdot2^{q-1}/q}$. The lines correspond to first order phase transitions between a high-entropy SYK-like phase and a low-entropy harmonic oscillator-like phase. They start at $(T=0, \mu_*)$ and end at the critical point $(T_c, \mu_c)$. At high temperatures, the system enters a supercritical phase.}
	\label{fig:fig2}
	\end{center}
\end{figure}

We report our results on the thermodynamic, transport, and chaos properties of the complex SYK model in the following sections. We show that there are two distinct phases in equilibrium, each leading to very different scalings of our observables. We then study the dependence of the diffusivity and conductivity matrices on chemical potential and temperature, in relation to the aforementioned phases. Lastly, we investigate a bound on diffusion imposed by the chaotic dynamics of the system. 

In order to emphasize that our methods are applicable to a range of parameters, we display results for different interaction orders $q$. To this extent, we fix the rescaled couplings $\J_{0,1}|_{\mu=0}=J_{0, 1}\sqrt{q2^{1-q}}=1$, which sets the results for different $q$ on equal footing and allows for a direct comparison to previously reported values for Majorana fermions~\cite{zanoci2022}. It also keeps $J$ independent of both $\mu$ and $\beta$. Furthermore, we only focus on the regime with $\mu\geq0$, since the sign of $\mu$ can be changed by simply swapping the roles of the creation and annihilation operators in our model. 

\subsection{Equilibrium phase diagram}
\label{sec:equilibrium_results}

We begin by investigating the phase diagram of the SYK model at finite chemical potential and temperature. In this model, a first order phase transition arises as a result of the competition between a high-entropy SYK-like phase and a low-entropy harmonic oscillator-like phase, and it has been extensively studied in the literature~\cite{azeyanagi2018,ferrari2019,cao2021thermodynamic,samui2021,tikhanovskaya2021a}.

\begin{figure}[tp]
	\begin{center}
	\includegraphics[width = \columnwidth]{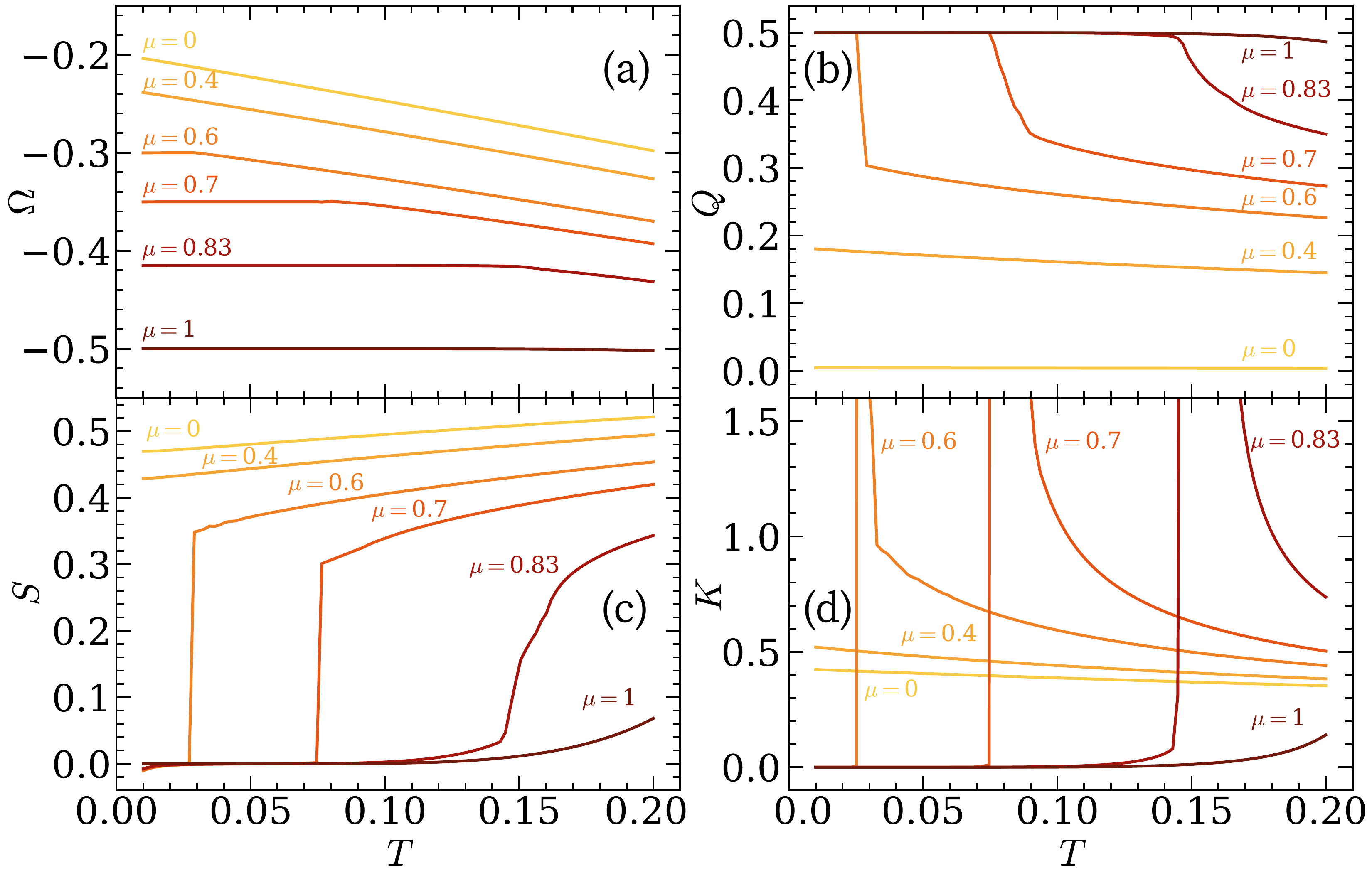}
	\caption{Phase transition characteristics of the SYK model for $q=4$. Grand canonical potential (a), charge (b), entropy (c), and charge compressibility (d) as a function of temperature. Signatures of a first order phase transition are seen for $0.52\leq\mu\leq0.83$. The thermodynamic derivatives $Q$ and $S$ experience a jump where the potential $\Omega$ changes slope, while the compressibility $K$ shows a discontinuity.}
	\label{fig:fig3}
	\end{center}
\end{figure}

The SD equations can have two distinct solutions depending on the values of $\mu$ and $\beta$. At small $\mu$, the behavior is similar to the Majorana SYK case. The model has a trivial perturbative expansion around the maximally mixed state at high temperatures, and a non-trivial conformal regime with an emergent approximate time-reparametriation symmetry at low temperatures~\cite{maldacena2016}. The latter regime also features a finite zero-temperature entropy and maximal chaos, reminiscent of nearly extremal black holes. Therefore, we label this region as the high-entropy or SYK-like phase~\cite{azeyanagi2018,ferrari2019}. 

On the other hand, in the limit of large $\mu$, the model behaves like a set of weakly coupled harmonic oscillators and the ground state is given by the unique Fock vacuum all the way to zero temperature ~\cite{azeyanagi2018,ferrari2019}. Hence the system is non-chaotic, has a vanishingly small entropy at low temperatures and an exponentially decaying Euclidean two-point function $G(\tau)\sim e^{-\mu\tau}$. We will refer to this as the low-entropy or harmonic oscillator-like phase~\cite{azeyanagi2018,ferrari2019}. 

\begin{figure}[tp]
	\begin{center}
	\includegraphics[width = \columnwidth]{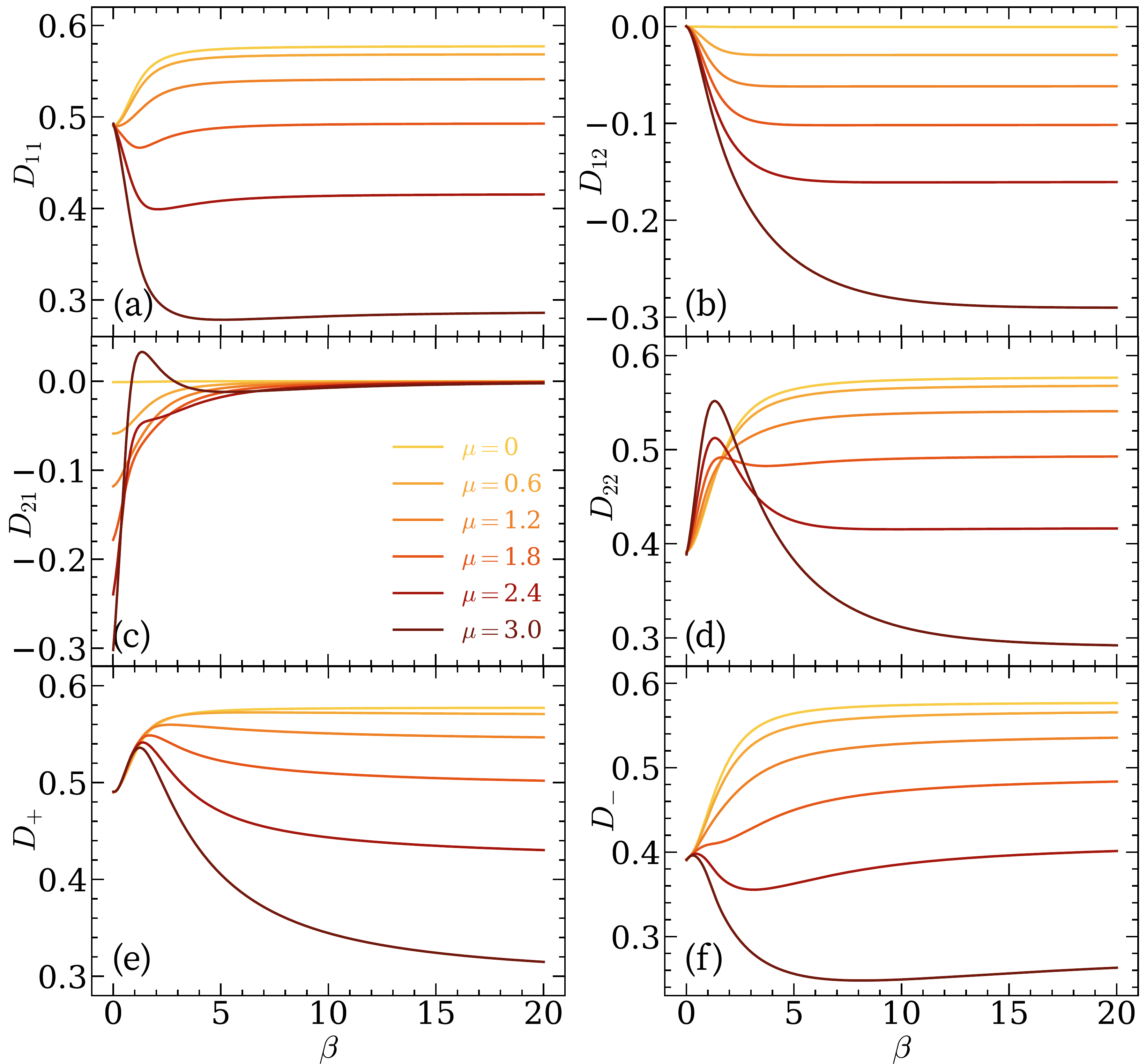}
	\caption{Temperature dependence of diffusivity $D$ for $q = 2$ and multiple values of $\mu$. Individual entries of the diffusivity matrix (a-d) and its eigenvalues (e-f) approach finite values at zero temperature in accordance with  \eqref{eq:diffusivity_q=2_zero}.}
	\label{fig:fig4}
	\end{center}
\end{figure}

The two solutions are separated by a finite first order phase transition line, which starts at $(T=0, \mu_*)$ and culminates at a critical point $(T_c, \mu_c)$ with asymmetric $q$-dependent critical exponents~\cite{azeyanagi2018,ferrari2019,cao2021thermodynamic}. For $q=4$, we have $\mu_*\approx0.52$, $T_c\approx0.16$, and $\mu_c\approx0.83$. At the critical point, the two solutions are identical and the transition becomes second order. For $T>T_c$, the SD equations have only one solution, corresponding to a high-temperature perturbative regime, and the system is in a supercritical phase~\cite{azeyanagi2018,ferrari2019}. The high- and low-entropy phases can be smoothly connected by going around the critical point, which emphasizes that there is no sharp distinction between them. We summarize these findings in the phase diagram of \figref{fig:fig2}. Note that all of our units are rescaled by a trivial factor of $J$ compared the diagrams in Refs.~\cite{azeyanagi2018,ferrari2019}. 

On the transition line, the values of $\Omega$ for the two solutions are equal and the two phases can coexist (see \figref{fig:fig3}(a)). Upon crossing the line, the Green's function jumps from one solution to the other, causing a discontinuity in the first order derivatives of the potential. This is illustrated in \figref{fig:fig3}(b-c), where the charge and entropy show clear signs of a first order phase transition for $\mu_*\leq\mu\leq\mu_c$. Consequently, second derivatives experience a singularity at the transition point, as exemplified by the charge compressibility $K$ in \figref{fig:fig3}(d). Notice that the $T=0$ state has a finite entropy and compressibility below $\mu_*$, while above $\mu_*$ it has maximal charge and zero entropy and compressibility. This is consistent with our previous description of the two phases.  

The same qualitative behavior is observed for all values of $q\geq 4$, with the transition line shrinking rapidly as $q$ increases (see \figref{fig:fig2}). We expect this transition to disappear completely in the infinite-$q$ limit, as can be seen explicitly from the thermodynamic potential in \eqref{eq:omega_large_q}. The low-entropy solution becomes favorable when the second term switches sign from negative to positive, which never happens at finite temperatures because $\tan\left(\frac{\pi v}{2}\right)>\frac{\pi v}{4}$ for all $v\in(0, 1)$. Analogously, there is no phase transition in the case of $q=2$ either. The grand canonical potential in \eqref{eq:omega_q=2} and its derivatives are smooth, continuous functions, and the Green's function always converges to its free-fermion value. 

\begin{figure}[tp]
	\begin{center}
	\includegraphics[width = \columnwidth]{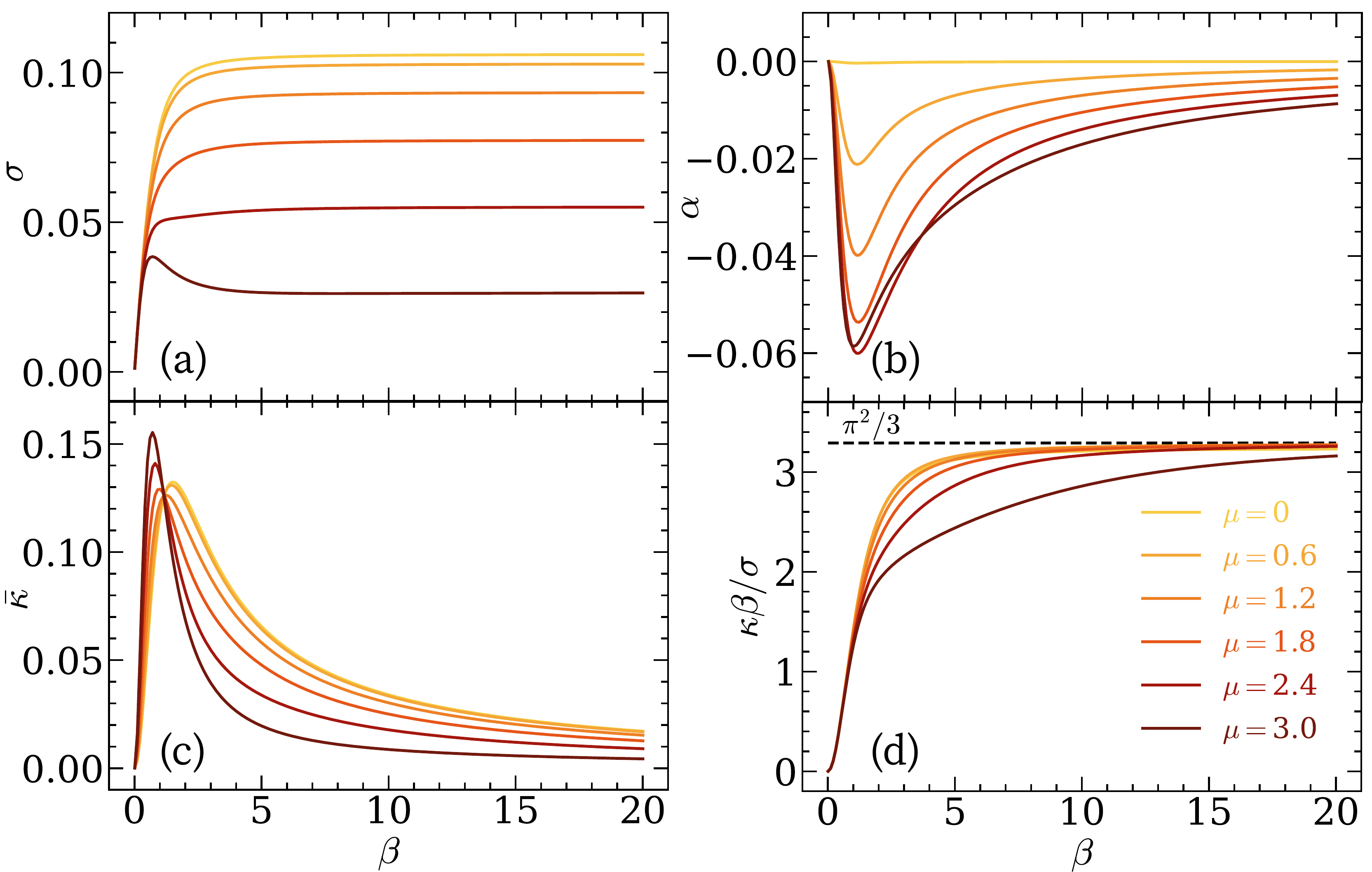}
	\caption{Temperature dependence of conductivity matrix $L$ for $q = 2$ and several values of $\mu$. Electrical conductivity (a) saturates to a constant at zero temperature, while thermoelectric (b) and thermal (c) conductivities approach zero as $1/\beta$. The $\mu$-dependence at low temperatures is quadratic, in agreement with \eqref{eq:conductivity_q=2_zero}. (d) The Wiedemann-Franz ratio converges to its free-fermion value of $\pi^2/3$.}
	\label{fig:fig5}
	\end{center}
\end{figure}

\subsection{Near-equilibrium transport}
\label{sec:nonequilibrium_results}

The presence of a phase transition has important consequences for both the transport coefficients and the Lyapunov exponents discussed next. In the high-entropy phase, we find that transport is diffusive and the Lyapunov exponent is non-vanishing. Since the high- and low-entropy phases are smoothly connected by going around the critical point via the super-critical phase, we expect that the dynamics is diffusive and chaotic throughout the phase diagram. However, we do observe extreme changes in the diffusivities and Lyapunov exponents in the vicinity of the phase transition line. Moreover, while these properties are expected to be non-vanishing, they can be very small and quite difficult to ascertain numerically. Therefore, when applicable, we will restrict our analysis to the SYK-like phase, where our quantities of interest are more straightforward to obtain. Lastly, for all the parameter regimes considered below, we checked numerically using the same open-system setup as in Ref.~\cite{zanoci2022}, that the NESS solutions of the full Kadanoff-Baym equations in the presence of weak driving indeed take on the form in \eqref{eq:ness_ansatz}. Thus our ansatz is justified. 

\begin{figure}[tp]
	\begin{center}
	\includegraphics[width = \columnwidth]{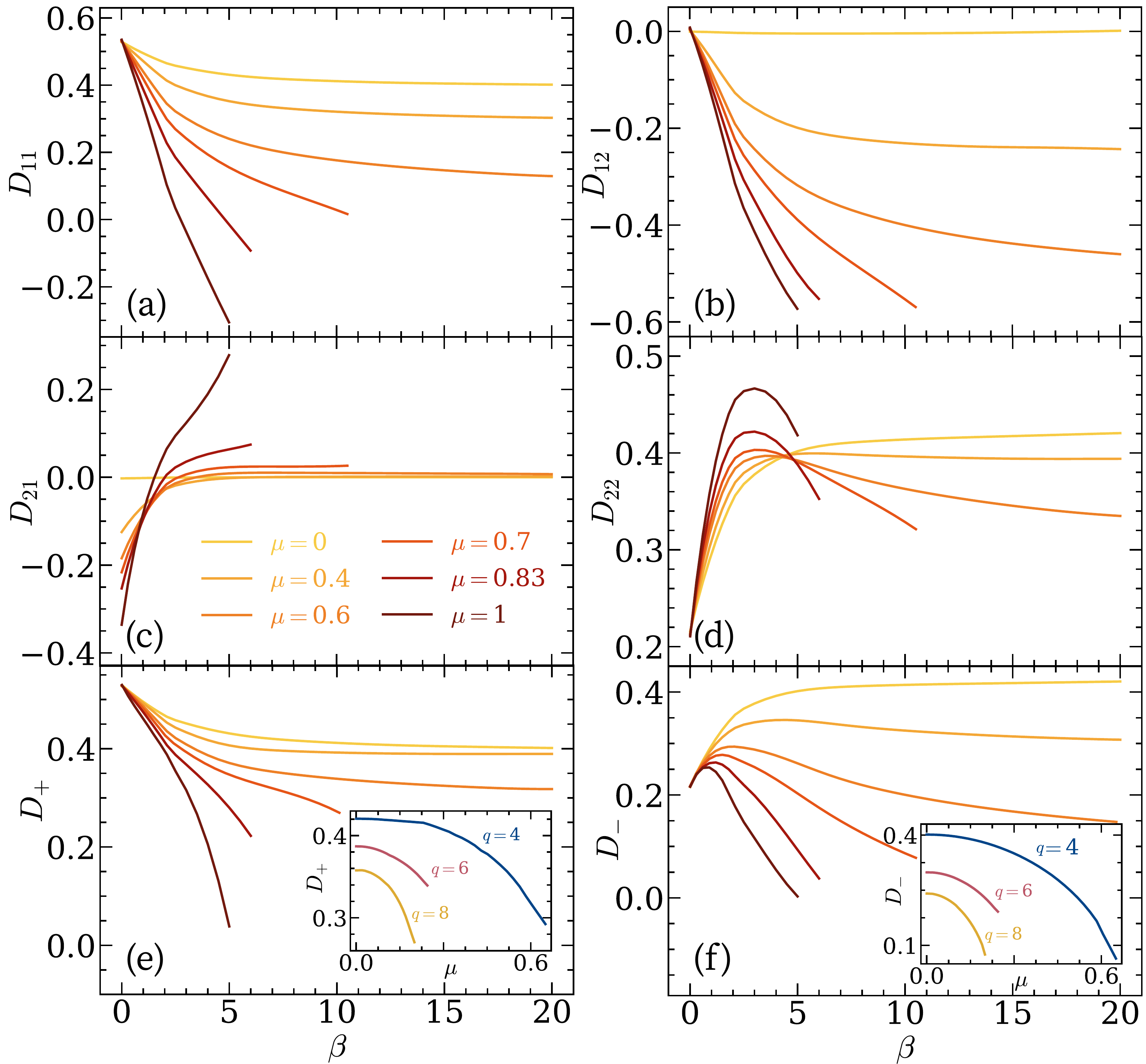}
	\caption{Temperature dependence of diffusivity $D$ for $q = 4$ and multiple values of $\mu$. The matrix elements $D_{ij}$ (a-d) and eigenvalues $D_{\pm}$ (e-f) are shown in the SYK-like phase. The eigenvalues reach a constant at low temperatures and its variation with $\mu$ for different $q$ is shown in the inset. The diffusivity decreases for both larger $\mu$ and $q$.}
	\label{fig:fig6}
	\end{center}
\end{figure}

We now proceed with our results for the simplest free-fermion case of $q=2$. We numerically compute all the integrals in \secref{sec:q=2_methods} and extract the transport coefficients. Their values are plotted in \figref{fig:fig4} and \figref{fig:fig5} as a function of inverse temperature and for different $\mu\in[0, 2J]$. In the limits of zero and infinite temperature, we were able to find the linear response functions analytically and obtained exact solutions for both $D$ and $L$ in \appref{sec:appendixC}. It is easy to check that they agree with our numerical results in \figref{fig:fig4} and \figref{fig:fig5} in the corresponding limits. We will elaborate below on the specific structure of the diffusivity and conductivity matrices in these limits. 

\begin{figure}[tp]
	\begin{center}
	\includegraphics[width = \columnwidth]{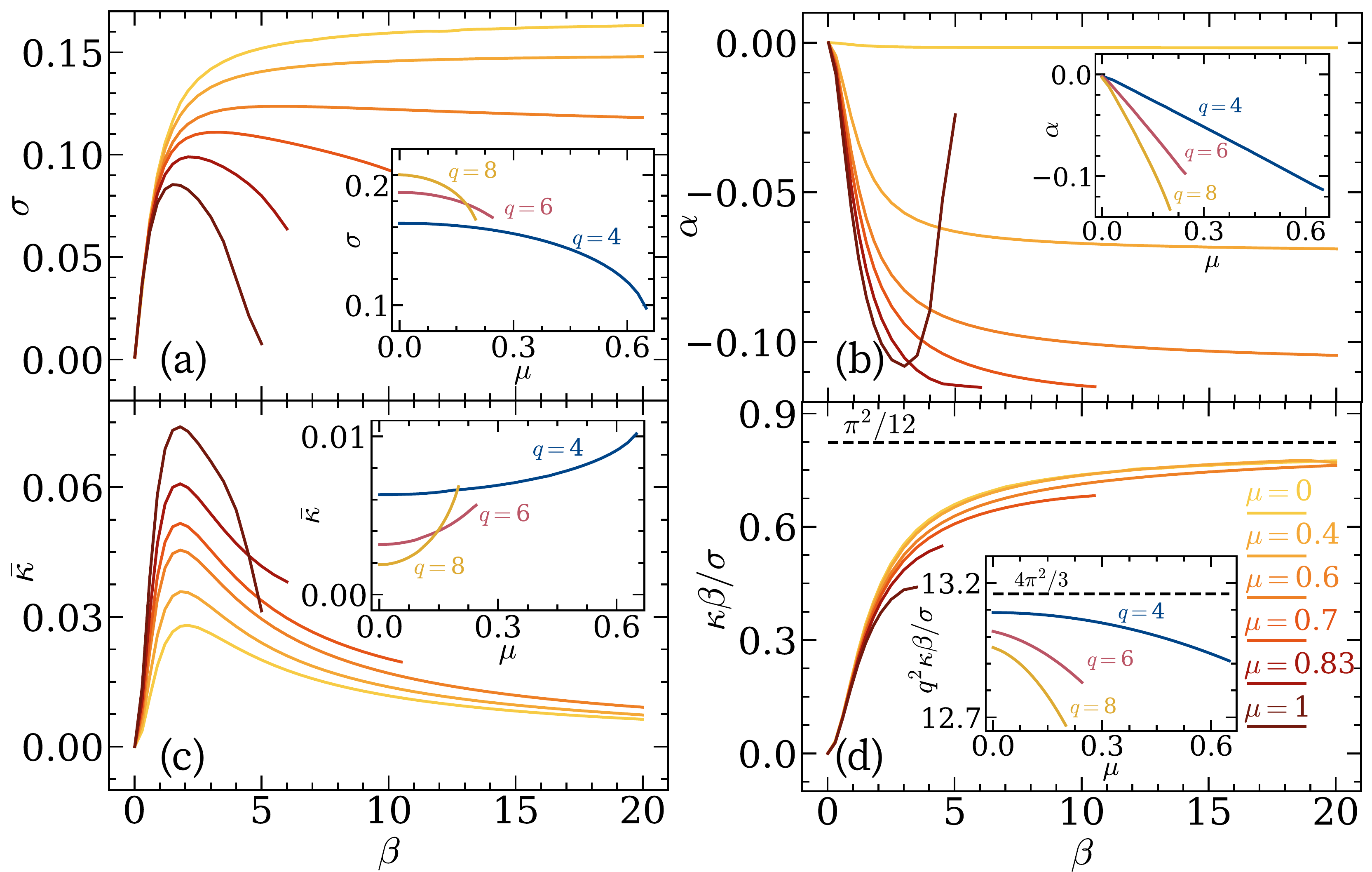}
	\caption{Temperature dependence of conductivity matrix $L$ for $q = 4$ and several values of $\mu$. Electrical (a), thermoelectric (b), and thermal (c) conductivities are shown in the SYK-like phase. The insets display their low-temperature asymptotic behavior as a function of $\mu$. (d) The Wiedemann-Franz ratio converges to $\pi^2/12$ for $q = 4$. The inset confirms that the scaling generalizes to $\kappa\beta/\sigma \to 4\pi^2/3q^2$ for other values of $q$.}
	\label{fig:fig7}
	\end{center}
\end{figure}

Next, we discuss our results for $q\geq4$. Since all these cases are very similar, we focus on $q=4$ in the main panels of \figref{fig:fig6} and \figref{fig:fig7}, with the understanding that the same conclusions hold for larger $q$. At small $\mu$, we recover the same behavior as in the Majorana case~\cite{zanoci2022}. For $\mu=0.7$ and $\mu = 0.83$, we encounter the phase transition within our range of temperatures, and the transport coefficients drop close to zero abruptly. At large $\mu$, we avoid the phase transition completely and directly enter the low-entropy phase. In this case, the conductivities and $D_{\pm}$ smoothly decrease as we lower the temperature. Notice that $D_{11}$ can become negative as we approach the low-entropy phase, which seems troubling at first. However, recall that only the eigenvalues $D_{\pm}$ are required to be positive to ensure the decay of charge and energy fluctuations, which we verify to be the case in \figref{fig:fig6}(e-f). 

There are a lot of features that emerge from the structure of the diffusivity matrix for both $q = 2$ (\figref{fig:fig4}) and $q\geq4$ (\figref{fig:fig6}). At high temperature, $D_{12} = 0$  and $D_{21}\sim \mu$. The eigenvalues $D_\pm$ are identical to the diagonal entries $D_{11,22}$ and approach finite $\mu$-independent values. At low temperature and away from the phase transition, the other off-diagonal entry vanishes $D_{21}=0$ and the eigenvalues $D_+=D_{22}$ and $D_-=D_{11}$ converge to $\beta$-independent values. The $\mu$-dependence of these numbers for different $q$ is shown in the inset of \figref{fig:fig6}(e-f) and in \eqref{eq:diffusivity_q=2_zero}. We see that the diffusivity decreases with both $\mu$ and $q$. These constraints on the diffusivity matrix and the generalized Einstein relations are enough to conclude that 
\begin{equation}
\begin{split}
    \sigma &= D_{11}K,\\
    \kappa &= D_{22}\gamma/\beta,
    \label{eq:sigma_and_kappa}
\end{split}
\end{equation}
at both high and low temperatures. These non-trivial relations are checked explicitly for $q=2$ in \appref{sec:appendixC}. The same dependence among transport coefficients was found for holographic theories and the SYK chain in the conformal limit~\cite{davison2017}. There it was attributed to the interplay between the global $U(1)$ charge and the emergent PSL$(2, {\rm I\!R})$ symmetry. It is interesting that here we see the same structure also emerge at infinite temperature.

The conductivity matrix can be examined in the same way (see \figref{fig:fig5} and \figref{fig:fig7}). At high temperature, all the conductivities are zero, while at low temperature, $\sigma$ is finite and $\kappa$ decays as $1/\beta$. In fact, as $T\to0$ we observe a linear-in-T resistivity $\sigma^{-1}$, above a background residual
resistivity $\sigma_0^{-1}$, according to the prediction in~\cite{guo2020,tikhanovskaya2021b} 

\begin{equation}
    \frac{1}{\sigma} = \frac{1}{\sigma_0}\left(1+4\alpha_G\frac{T}{J}\right),
    \label{eq:linear_in_T}
\end{equation}
where $\alpha_G$ is a known numerical constant~\cite{maldacena2016}. A linear fit to our data yields $\alpha_G\approx0.194$ for $\mu=0$ and $q=4$, which is very close to the literature value $\alpha_G\approx0.187$ (e.g. Fig. 9 in \cite{maldacena2016}). This linear-in-$T$ resistivity is a common feature of many non-Fermi liquid models~\cite{parcollet1999,song2017,patel2018,chowdhury2018,guo2020,chowdhury2021sachdev}. The $\mu$-dependence of the conductivities at low temperature is available in the inset of \figref{fig:fig7} and in \eqref{eq:conductivity_q=2_zero}. We notice that $\alpha$ scales linearly with $\mu$, while $\sigma$ and $\bar{\kappa}$ have a dependence that is closer to quadratic. We can also combine these results to show that our model has a non-vanishing thermopower all the way to zero temperature, as discussed further in \appref{sec:appendixD}

These observations about the structure of the conductivity matrix lead us to believe that the Wiedemann-Franz ratio $\kappa\beta/\sigma$ approaches a constant at zero temperature. Indeed we find numerically in \figref{fig:fig7}(d) for $q=4$ and analytically from \eqref{eq:conductivity_q=2_zero} for $q=2$ that 

\begin{equation}
    \lim_{\beta\to\infty}\frac{\kappa\beta}{\sigma} = \frac{4\pi^2}{3q^2},
    \label{eq:WF_ratio}
\end{equation}
in agreement with the results of Ref.~\cite{davison2017}. This also holds for other values of $q$, as long as we are still in the SYK-like phase, as shown in the inset of \figref{fig:fig7}(d). The slight deviations at larger values of $\mu$ are caused by our inability to numerically reach low enough temperatures without crossing the phase transition. At zero temperature, we can combine the two results above to find that the ratio of diffusivities obeys~\cite{davison2017} 

\begin{equation}
    \frac{D+}{D_-} = \frac{D_{22}}{D_{11}} = \frac{4\pi^2}{3q^2}\frac{K}{\gamma}.
    \label{eq:diffusivity_ratio}
\end{equation}

Finally, we are ready to present our findings in the large $q$ limit, following the derivation in \secref{sec:large_q_methods} and \appref{sec:appendixC}. We find that $D_{12}=0$, which together with the Einstein relations, is enough to conclude that  \eqref{eq:sigma_and_kappa} holds for all values of $\mu$ and $\beta$. Moreover, we can combine the conductivities in \eqref{eq:sigma_large_q} and \eqref{eq:kappa_large_q} to arrive at the Wiedemann-Franz ratio

\begin{equation}
    \frac{\kappa\beta}{\sigma} = \frac{4\pi^2v^2}{3q^2}.
\end{equation}
At zero temperature, $v\to1$ and we recover the results in \eqref{eq:WF_ratio} and \eqref{eq:diffusivity_ratio}. Therefore, all the previously found features of transport at finite $q$ are also applicable to the infinite $q$ regime. In addition, this expansion provides compact solutions for all the transport coefficients over the entire parameter range (see \appref{sec:appendixC}).

\begin{figure}[tp]
	\begin{center}
	\includegraphics[width = \columnwidth]{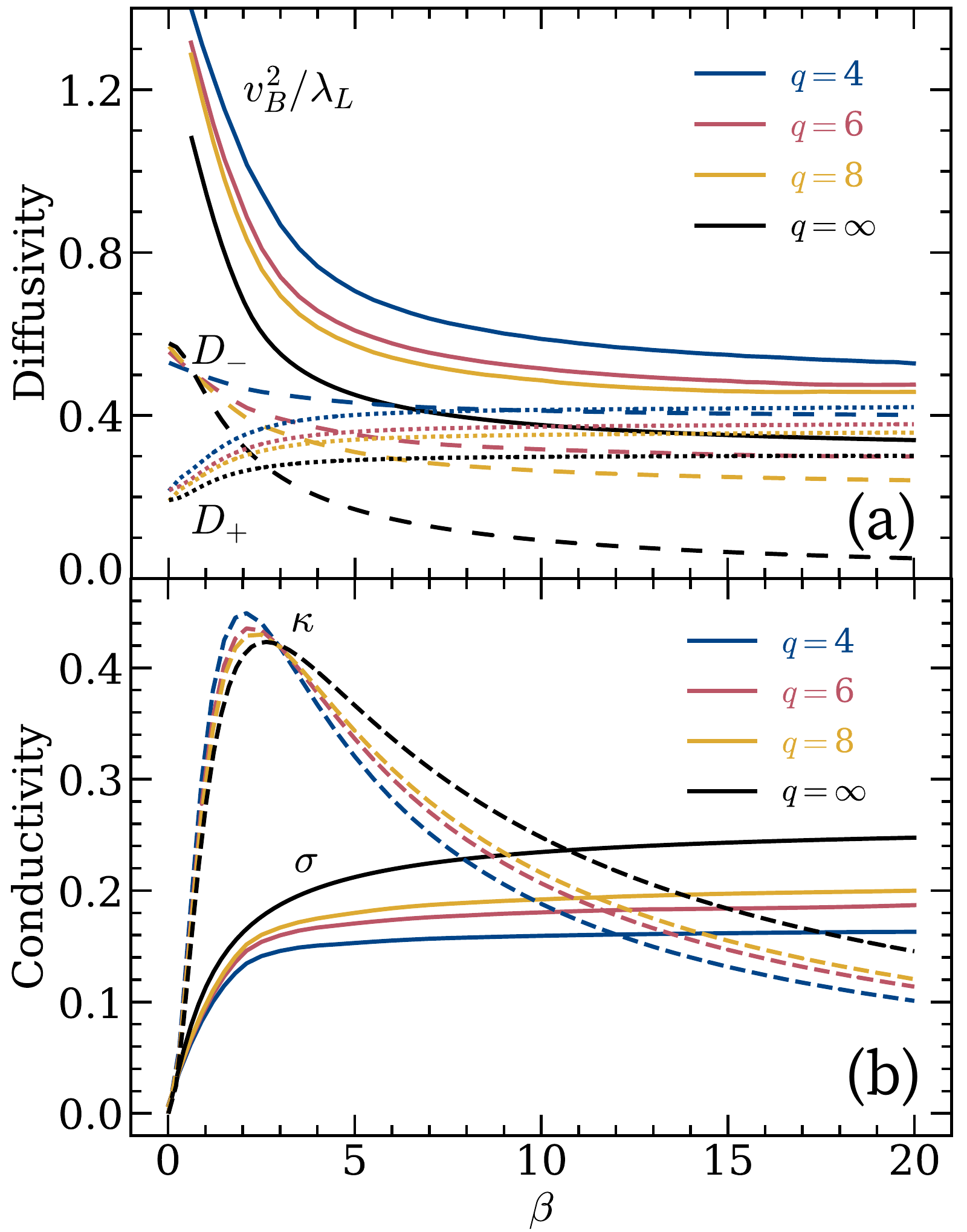}
	\caption{Comparison between finite and infinite $q$ diffusivities and conductivities at $\mu = 0$. (a) The diffusivity eigenvalues $D_{\pm}$ (dashed and dotted lines) are upper bounded by the chaos propagation rate $v_B^2/\lambda_L$ (solid lines) for all $q$. (b) The electrical conductivity $\sigma$ (solid lines) and thermal conductivity $\kappa$ (dashed lines) show little variation with $q$.}
	\label{fig:fig8}
	\end{center}
\end{figure}

In order to make a fair comparison to the finite $q$ results, we restrict ourselves to the case $\mu=0$, for reasons explained in \appref{sec:appendixC}. The charge and energy diffusion modes decouple and we are left with a diagonal diffusivity matrix 

\begin{align}
    D_- &= D_{11} = \frac{\J_1^2}{\J}\cos(\frac{\pi v}{2}),\\
    D_+ &= D_{22} = \frac{\J_1^2}{3\J}\left(\frac{\pi v}{2}\sin(\frac{\pi v}{2})+\cos(\frac{\pi v}{2})\right).
    \label{eq:D_+_large_q}
\end{align}
The temperature dependence enters the expressions implicitly through $v$ (see \eqref{eq:v}). We plot these results in \figref{fig:fig8}(a), together with the finite $q$ values obtained numerically by following the prescription in \secref{sec:transport_methods}. All the curves obey the same pattern and the agreement with the $q\to\infty$ result clearly improves with increasing $q$. The energy diffusion constant $D_{22}$ agrees with our previous answer for Majorana fermions~\cite{zanoci2022}, as expected for $\mu=0$. Similarly, the electrical and thermal conductivities are given by 

\begin{align}
    \sigma &= \frac{\J_1^2}{\J^2}\frac{\pi v}{4}, \\
    \kappa &= \frac{\J_1^2}{\beta q^2 \J^2}\frac{(\pi v)^3}{3}. 
\end{align}
These are shown in \figref{fig:fig8}(b) next to their finite $q$ counterparts. The agreement is quite good even for moderate values of $q$.

\subsection{Chaos}
\label{sec:chaos_results}

In this section, we further explore the connections between transport and many-body chaos. We numerically diagonalize the kernel introduced in \secref{sec:chaos_methods} and extract the Lyapunov exponent $\lambda_L$ and the butterfly velocity $v_B$. Both of these quantities only weakly depend on the chemical potential $\mu$ in the SYK-like phase. In particular, we checked that in the limit of infinite temperature and finite charge density, both $\lambda_L$ and $v_B$ saturate the bounds proposed in Ref.~\cite{chen2020}. Upon approaching the phase transition, they decay exponentially with $\mu$ and tend to zero in the low-entropy phase~\cite{Bhattacharya2017,Sorokhaibam2020}. This is not surprising, since the conserved $U(1)$ charge constrains the phase-space dynamics of the system. A large chemical potential eventually renders the system integrable, as manifested by a transition to the harmonic oscillator-like phase, for which very weak chaotic behavior is expected.

\begin{figure}[tp]
	\begin{center}
	\includegraphics[width = \columnwidth]{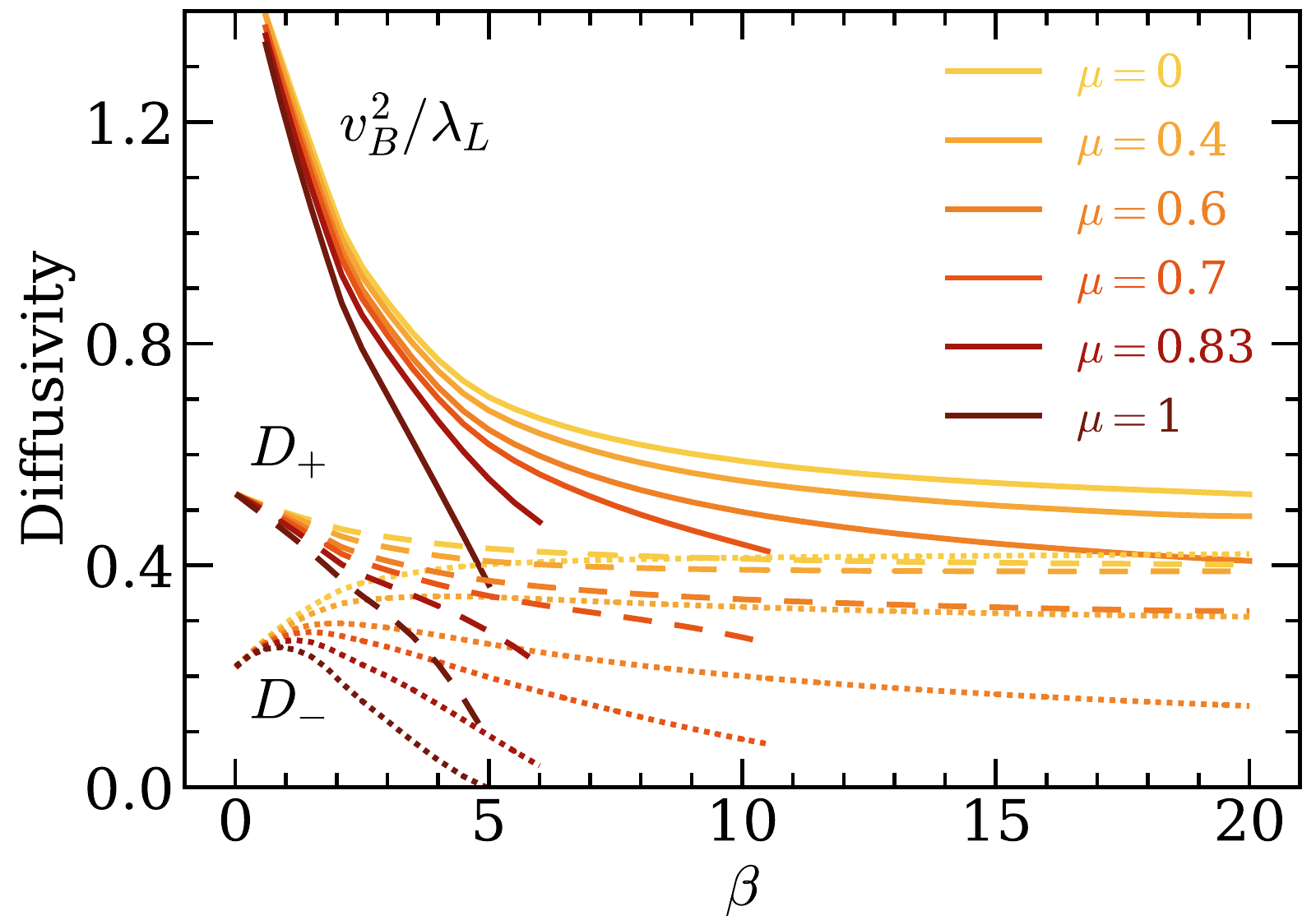}
	\caption{The diffusivity eigenvalues $D_{\pm}$ (dashed and dotted lines) and the chaos bound $v_B^2/\lambda_L$ (solid lines) for $q = 4$ and different chemical potentials. The bound is saturated in the conformal limit. All three quantities drop to zero outside the SYK-like phase.}
	\label{fig:fig9}
	\end{center}
\end{figure}

The ratio $v_B^2/\lambda_L$ exhibits a similar behavior. In \figref{fig:fig9} we plot its temperature dependence in the SYK-like phase for $q=4$ and compare it to the diffusivity eigenvalues $D_\pm$, since these are more physically relevant than the diagonal entries of $D$. Our results indicate that $D_\pm\leq v_B^2/\lambda_L$ at all temperatures, suggesting that chaos upper bounds diffusion. This inequality generalizes the bound we previously found for energy diffusion in Majorana SYK chains~\cite{zanoci2022}. We observe that in the limit of zero temperature and for $\mu<\mu_*$, the thermal diffusivity saturates the chaos bound $D_{22}=D_+=v_B^2/\lambda_L$. This remarkable result is a consequence of the fact that the same reparameterization degrees of freedom are responsible both for thermal diffusion and the OTOC chaos dynamics~\cite{maldacena2016,gu2017diffusion}. The charge diffusivity, on the other hand, is not easily related to chaos in this model~\cite{davison2017}. In \figref{fig:fig8}(a) we verify that the same results hold for other values of $q$, as well as in the large $q$ limit.


The fact that the SYK chain reaches this equality in the conformal limit has been previously shown for both Majorana and complex fermions~\cite{gu2017diffusion,davison2017}. However, our method for calculating the diffusivities at arbitrary $\mu$ and $\beta$ allows us to confirm the inequality $D_\pm\leq v_B^2/\lambda_L$ beyond the conformal or large $q$ limits~\cite{choi2021}. A similar bound has been found for other families of models as well~\cite{gu2017,lucas2016,chen2020}. We should mention that this inequality is by no means universal, since there are examples of theories where it holds in the opposite direction~\cite{blake2016_1,blake2016_2,blake2017}. A more rigorous upper bound on diffusivity can be written in the form of $D\leq v^2\tau_{\text{eq}}$~\cite{hartman2017,lucas2017constraints,han2018,hartnoll2021planckian}, where $v\sim v_B$ is the operator growth velocity and $\tau_{\text{eq}}$ is the local equilibration timescale, which can be much larger than the Lyapunov timescale $1/\lambda_L$~\cite{hartman2017}.

\section{Discussion}
\label{sec:discussion}

This work has described the thermodynamic, transport, and chaos properties of an SYK chain with general $q$-body interactions. Our main result is a detailed analysis of the near-equilibrium response of local Green's functions to small external biases. More specifically, we expanded the Green's function of each SYK cluster to first order in the non-equilibrium corrections $F_{\mu,\beta}^\gtrless$ due to constant chemical potential and temperature gradients. We were then able to express all the conserved charges and their associated currents in terms of these functions. The calculations were carried out analytically for $q=2$ and $q\to\infty$, and numerically for all other values of $q$ using the solutions of the Schwinger-Dyson equation described in \appref{sec:appendixA}. This allowed us to fully characterize the mixed thermoelectric response of the model in terms of its diffusivity, conductivity, and susceptibility matrices. Moreover, we showed that the eigenvalues of the diffusivity matrix satisfy the inequality $D_{\pm}\leq v_B^2/\lambda_L$ at all temperatures, with equality achieved for $D_+$ in the conformal limit. This result generalizes our previous bound on energy diffusion in the case of Majorana fermions~\cite{zanoci2022} and establishes a connection between transport and chaos in the SYK model.

Our analysis has revealed new features in the structure of the transport coefficients. In particular, we showed that one of the off-diagonal entries of the diffusivity matrix approaches zero at both high and low temperatures, as well as in the large $q$ limit. Together with the Einstein relations, this results in a simplified expression for the conductivities in \eqref{eq:sigma_and_kappa}, which was previously established only for SYK and holographic models in the conformal limit~\cite{davison2017}. Additionally, we showed that the Wiedemann-Franz ratio approaches the finite value $4\pi^2/3q^2$ at zero temperature, in agreement with Ref.~\cite{davison2017}. For $q=2$ we recover the universal Fermi liquid prediction in the form of the Lorenz number $\pi^2/3$. We should emphasize that for $q\geq 4$, this result is not universal and depends on the specific choice of interaction between clusters~\cite{davison2017}.

Although our methods are valid for arbitrary values of $\beta$ and $\mu$, we have to be careful when interpreting our results close to the phase transition between the high- and low-entropy phases and in the low-entropy phase. Specifically, various transport quantities experience a sudden drop or divergence when crossing the transition. Moreover, since the low-entropy phase has very small values of the diffusivity and Lyapunov exponent, we cannot always robustly study the low temperature limit of our observables past the phase transition. Hence, in some instances, we have to restrict ourselves to small values of $\mu$, where the SYK-like phase extends all the way to zero temperature. Given the weakly coupled nature of the low-entropy phase, other analytical methods may be useful in that regime, if the physics is of interest.

Our work paves the way for further analytical and numerical studies of linear response in quantum many-body systems. In this paper, we focused on the zero-frequency response of a uniform one-dimensional chain, but generalizations should be straightforward. For example, the frequency dependence of the transport coefficients can be extracted by imposing a time-dependent oscillatory bias~\cite{zanoci2022,kuhlenkamp2020}. Our methods are also suited for other higher-dimensional non-Fermi liquid models built from SYK clusters~\cite{chowdhury2018,patel2018}, or more general theories with tractable local Green's functions. The same ideas can in principle be applied to study transport in more conventional spin systems~\cite{bertini2021,weimer2021,landi2021,zanoci2021}, where the NESS is approximated as a tensor network, although the details of this calculation are more complicated. 

Despite its success predicting the main features of transport, linear response theory has some limitations as a probe of non-equilibrium dynamics in quantum systems. It would be interesting to investigate the effect of strong driving on our SYK system, where non-linear effects, such as Joule heating, play an important role. To capture the physics beyond linear response, one would have to solve the full Kadanoff-Baym for the system out-of-equilibrium~\cite{zanoci2022}. Probing non-linear transport and out-of-equilibrium phase transitions are both interesting future directions.

The SYK chain discussed in this paper is a solvable theoretical model displaying some of the major properties of a non-Fermi liquid~\cite{chowdhury2021sachdev}. However, its non-Fermi liquid behavior is yet to be observed experimentally. In recent years, multiple experimental realizations~\cite{franz2018,rahmani2019,danshita2017,wei2021,pikulin2017,chew2017,yang2018,chen2018} and quantum simulations~\cite{garcia2017,luo2019,babbush2019,behrends2022} of SYK have been proposed. These include ultracold atom experiments~\cite{danshita2017,wei2021}, Majorana modes at the interface of a topological insulator and superconductor~\cite{pikulin2017}, semiconductor wires coupled through a disordered quantum dot~\cite{chew2017}, superconducting circuits~\cite{yang2018}, and graphene flakes~\cite{chen2018}. The latter configuration, based on the zeroth Landau level in graphene flakes with irregular boundaries subject to strong magnetic fields, is especially well suited for probing mesoscopic transport in the complex SYK model~\cite{can2019,kruchkov2020}. One could use this setup to look for signatures of a linear-in-$T$ resistivity at low temperatures according to \eqref{eq:linear_in_T}. It has also been suggested that measurements of the thermopower can serve as an indicator of the non-vanishing residual entropy at low temperatures~\cite{kruchkov2020}. This opens up the possibility of directly comparing our theoretical predictions with actual experimental data. 

\begin{acknowledgments}
We are grateful to Nikolay Gnezdilov, Aavishkar Patel, Nilakash Sorokhaibam, and Maria Tikhanovskaya for helpful discussions related to the numerical solutions of the SYK model. C.Z. acknowledges financial support from the Harvard-MIT Center for Ultracold Atoms through NSF Grant No. PHY-1734011. The work of B.S. is supported in part by AFOSR grant FA9550-19-1-0360.
\end{acknowledgments}

\bibliographystyle{apsrev4-2}
\bibliography{references}

\appendix

\section{Numerical solutions of the SD equations}
\label{sec:appendixA}

As mentioned in the main text, the Schwinger-Dyson equations in both real and imaginary time are solved numerically for the case of a single isolated SYK cluster in equilibrium. Our approach in Euclidean time is almost identical to the original one for Majorana fermions~\cite{maldacena2016}:

\begin{enumerate}
    \item Initialize $G(i\omega_n)$ with the free-fermion propagator $(i\omega_n+\mu)^{-1}$ and compute its inverse Fourier transform $G(\tau)$. 
    \item Calculate $\Sigma(\tau)$ using the second line in \eqref{eq:system_SD_2} and Fourier transform it to $\Sigma(i\omega_n)$.
    \item Compute a new Green's function $\tilde{G}(i\omega_n)$ from the first line in \eqref{eq:system_SD_2} and get its inverse Fourier transform $\tilde{G}(\tau)$. 
    \item Perform a weighted update $G(\tau)\leftarrow (1-\alpha)G(\tau)+\alpha\tilde{G}(\tau)$ with $\alpha = 0.3$.
    \item Repeat steps 2-4 until the iterative procedure converges $\max_\tau|G(\tau)-\tilde{G}(\tau)|<\epsilon$ with $\epsilon=10^{-12}$.
\end{enumerate}
The imaginary time domain is discretized as $\tau_n=nd\tau$, where $d\tau = \beta/M$, $-M\leq n\leq M$, and $M=2^{20}$. Similarly, the Matsubara frequencies are given by  $\omega_n = (2n+1)d\omega$, with $d\omega = \pi/\beta$ and $-M\leq n\leq M-1$. All the thermodynamic properties are then derived from the grand canonical potential in \eqref{eq:grand_canonical}.

Our algorithm for obtaining the real-time Green's functions is an extension of the method used in Ref.~\cite{eberlein2017} for Majorana fermions:

\begin{enumerate}
    \item Initialize $G^\gtrless(t)$ with the $q=2$ result in \eqref{eq:green_q=2}.
    \item Calculate $\Sigma^\gtrless(t)$ using the second line in \eqref{eq:SD_real_time} (for a single cluster $G_x^\gtrless=G_{x\pm1}^\gtrless=G^\gtrless$). Evaluate the retarded self-energy $\Sigma^R(t) = \Theta(t)(\Sigma^>(t)-\Sigma^<(t))$ and its Fourier transform $\Sigma^R(\omega)$.
    \item Compute $G^R(\omega)$ from the first line in \eqref{eq:system_SD_2} and find the spectral function $A(\omega)=-2\Im G^R(\omega)$.
    \item Determine a new Green's function $\tilde{G}^\gtrless(\omega)$ from the FDT in \eqref{eq:fdt_2} and get its inverse Fourier transform $\tilde{G}^\gtrless(t)$.
    \item Perform a weighted update $G^\gtrless(t)\leftarrow (1-\alpha)G^\gtrless(t)+\alpha\tilde{G}^\gtrless(t)$ with $\alpha = 0.7$.
    \item Repeat steps 2-5 until the iterative procedure converges $\max_t|G^\gtrless(t)-\tilde{G}^\gtrless(t)|<\epsilon$ with $\epsilon=10^{-5}$.
\end{enumerate}
The real time domain is discretized as $t_n=ndt$, where $dt = 0.05$, $-M\leq n\leq M$, and $M=10^4$. Similarly, Fourier transform frequencies are given by $\omega_n = nd\omega$, with $d\omega = 2\pi/Mdt$ and $-M\leq n\leq M$. At large values of $\mu$ and $\beta$, the above procedure can experience convergence issues. In order to mitigate this problem, we perform an additional annealing step, where we start at a high temperature and gradually lower it while re-running the algorithm with the Green's functions initialized to the previously converged values from a higher temperature run. 

\section{Charge and energy currents}
\label{sec:appendixB}

In this section, we present a derivation of the formulas for the charge and energy currents introduced in \secref{sec:transport_methods}. The non-equilibrium expectation value of any operator can be computed in the Keldysh formalism using the generating functional~\cite{kamenev2011field,stefanucci2013nonequilibrium}. For a one-dimensional chain, the charge current at site $x$ is given by 

\begin{equation}
\begin{split}
    j_x^Q &= i[Q_x, H^{x, x+1}] = i[Q_x, H_1^{x, x+1}] \\
    &\hspace{-0.2cm}= i\sum_{\{i\},\{j\}}\left(J_{i_1\ldots j_{\frac{q}{2}}}^{(1)} (c_{i_1}^{x})^\dagger\cdots (c_{i_{\frac{q}{2}}}^x)^\dagger c_{j_1}^{x+1}\cdots c_{j_{\frac{q}{2}}}^{x+1}-\hc\right),
\end{split}
\end{equation}
and its expectation value is similar to the on-bond energy 

\begin{equation}
\begin{split}
    \langle j_x^Q\rangle &= -\frac{J_1^2}{2}\int_{-\infty}^{t}\diff t_1 \Big(G_x^>(t, t_1)^{\frac{q}{2}} G_{x+1}^<(t_1, t)^{\frac{q}{2}} \\
    &- G_x^<(t_1, t)^{\frac{q}{2}} G_{x+1}^>(t, t_1)^{\frac{q}{2}} + \hc \Big).
    \label{eq:j_Q}
\end{split}
\end{equation}
This result is equivalent to the tunneling current formula derived in Ref.~\cite{cheipesh2020} for $q=2$.

In a similar fashion, the energy current at site $x$ is 

\begin{equation}
\begin{split}
    j_x^E &= i[H^{x-1, x}, H^{x, x+1}]\\
    &= i\left(\frac{1}{2}[H_0^x, H_1^{x, x+1}-H_1^{x-1, x}] + [H_1^{x-1, x},H_1^{x, x+1}]\right).
\end{split}
\end{equation}
The calculation of its expectation value mirrors the one for the energy current in the Majorana case~\cite{zanoci2022}, except that now we have extra contributions form the Hermitian conjugate terms in the interaction Hamiltonian

\begin{equation}
    \langle j_x^E\rangle = \frac{1}{2}J_1^2J^2\Re(j_{++}^{x-1, x, x+1}+j_{+-}^{x-1, x, x+1}),
    \label{eq:j_E}
\end{equation}

\begin{widetext}
\begin{align}
    j_{++}^{x-1, x, x+1} &= \int_{-\infty}^{t}\diff t_2 \int_{-\infty}^{t_2} \diff t_1 G_{x-1}^<(t_1, t)^{\frac{q}{2}}G_{x}^<(t_2, t)^{\frac{q}{2}-1}G_{x}^>(t, t_1)^{\frac{q}{2}-1}G_{x}^>(t_2, t_1)G_{x+1}^>(t, t_2)^{\frac{q}{2}}\nonumber \\
    &- \int_{-\infty}^{t}\diff t_2 \int_{-\infty}^{t_2} \diff t_1 G_{x-1}^>(t, t_1)^{\frac{q}{2}}G_{x}^>(t, t_2)^{\frac{q}{2}-1}G_{x}^<(t_1, t)^{\frac{q}{2}-1}G_{x}^<(t_1, t_2)G_{x+1}^<(t_2, t)^{\frac{q}{2}}\nonumber \\
    &+ \int_{-\infty}^{t}\diff t_2 \int_{-\infty}^{t_2} \diff t_1 G_{x-1}^<(t_2, t)^{\frac{q}{2}}G_{x}^<(t_1, t)^{\frac{q}{2}-1}G_{x}^>(t, t_2)^{\frac{q}{2}-1}G_{x}^<(t_1, t_2)G_{x+1}^>(t, t_1)^{\frac{q}{2}}\nonumber \\
    &- \int_{-\infty}^{t}\diff t_2 \int_{-\infty}^{t_2} \diff t_1 G_{x-1}^>(t, t_2)^{\frac{q}{2}}G_{x}^>(t, t_1)^{\frac{q}{2}-1}G_{x}^<(t_2, t)^{\frac{q}{2}-1}G_{x}^>(t_2, t_1)G_{x+1}^<(t_1, t)^{\frac{q}{2}}, \label{eq:j_++}\\
    j_{+-}^{x-1, x, x+1} &= \int_{-\infty}^{t}\diff t_2 \int_{-\infty}^{t_2} \diff t_1 G_{x-1}^<(t_1, t)^{\frac{q}{2}}G_{x}^>(t_2, t)^{\frac{q}{2}-1}G_{x}^>(t, t_1)^{\frac{q}{2}-1}G_{x}^>(t_2, t_1)G_{x+1}^<(t, t_2)^{\frac{q}{2}}\nonumber \\
    &- \int_{-\infty}^{t}\diff t_2 \int_{-\infty}^{t_2} \diff t_1 G_{x-1}^>(t, t_1)^{\frac{q}{2}}G_{x}^<(t, t_2)^{\frac{q}{2}-1}G_{x}^<(t_1, t)^{\frac{q}{2}-1}G_{x}^<(t_1, t_2)G_{x+1}^>(t_2, t)^{\frac{q}{2}}\nonumber \\
    &+ \int_{-\infty}^{t}\diff t_2 \int_{-\infty}^{t_2} \diff t_1 G_{x-1}^>(t_2, t)^{\frac{q}{2}}G_{x}^<(t_1, t)^{\frac{q}{2}-1}G_{x}^<(t, t_2)^{\frac{q}{2}-1}G_{x}^<(t_1, t_2)G_{x+1}^>(t, t_1)^{\frac{q}{2}}\nonumber \\
    &- \int_{-\infty}^{t}\diff t_2 \int_{-\infty}^{t_2} \diff t_1 G_{x-1}^<(t, t_2)^{\frac{q}{2}}G_{x}^>(t, t_1)^{\frac{q}{2}-1}G_{x}^>(t_2, t)^{\frac{q}{2}-1}G_{x}^>(t_2, t_1)G_{x+1}^<(t_1, t)^{\frac{q}{2}}.
    \label{eq:j_+-}
\end{align}
\end{widetext}
These expressions simplify greatly for a uniform chain in the near-equilibrium linear response regime, as shown in \secref{sec:transport_methods}.

\section{Exact calculations of the transport coefficients}
\label{sec:appendixC}

In \secref{sec:transport_methods} we introduced a simple non-equilibrium correction $F_{\mu,\beta}^\gtrless(t)$ to the Green's functions in the linear response regime. We showed that the conserved quantities and their currents can be expanded to first order in this function. Next, we will consider special cases where it is possible to analytically compute the equilibrium Green's functions, and hence also $F_{\mu,\beta}^\gtrless(t)$. In particular, we will provide detailed derivations for the susceptibility, diffusivity, and conductivity matrices in the limit of zero and infinite temperature for $q=2$, as well as in the $q\to\infty$ limit at arbitrary temperature. This will be a continuation of our discussion of these limits in \secref{sec:q=2_methods} and \secref{sec:large_q_methods}.

\subsection{\texorpdfstring{$q=2$}{q=2} limit}

The SYK Hamiltonian for $q=2$ is equivalent to a random hopping model. The partition function can be computed directly from the free-fermion picture~\cite{maldacena2016}. After fixing the reference energy level to match our convention for the charge in \eqref{eq:charge}, we have 

\begin{equation}
    Z = \prod_{|\omega|<2J} e^{-\beta\mu/2} \left(1+e^{-\beta(\omega-\mu)}\right)
\end{equation}
Note that unlike Majoranas, complex fermions are not paired up when performing the product over all modes. The thermodynamic potential becomes

\begin{equation}
    \Omega = \int_{-2J}^{2J} \diff\omega g(\omega) \left(\log(1+e^{-\beta(\omega-\mu)})-\frac{\beta\mu}{2}\right),
    \label{eq:omega_q=2}
\end{equation}
where we introduced the normalized density of states $g(\omega) = A(\omega)/2\pi$. The entries of the susceptibility matrix can be found by taking second derivatives of the potential 

\begin{align}
    \chi_{11} &= \int_{-2J}^{2J} \diff\omega \frac{\beta}{4\pi J\cosh^2(\beta(\omega-\mu)/2)}\sqrt{1-\left(\frac{\omega}{2J}\right)^2}, \\
    \chi_{12} &= \int_{-2J}^{2J} \diff\omega \frac{\beta^2(\omega-\mu)}{4\pi J\cosh^2(\beta(\omega-\mu)/2)}\sqrt{1-\left(\frac{\omega}{2J}\right)^2}, \\
    \chi_{22} &= \int_{-2J}^{2J} \diff\omega \frac{\beta^2(\omega-\mu)^2}{4\pi J\cosh^2(\beta(\omega-\mu)/2)}\sqrt{1-\left(\frac{\omega}{2J}\right)^2}. 
\end{align}

Now switching over to transport, from \eqref{eq:green_q=2} we can deduce 

\begin{align}
    F_\beta^>(t) &= \int_{-2J}^{2J} \diff\omega \frac{i(\mu-\omega)e^{-i\omega t}\nabla\beta}{4\pi J\cosh^2(\beta(\omega-\mu)/2)}\sqrt{1-\left(\frac{\omega}{2J}\right)^2},\\
    F_\mu^>(t) &= \int_{-2J}^{2J} \diff\omega \frac{i\beta e^{-i\omega t}\nabla\mu}{4\pi J\cosh^2(\beta(\omega-\mu)/2)}\sqrt{1-\left(\frac{\omega}{2J}\right)^2}.
\end{align}
All these integrals can always be performed numerically, but in the special case of high or low temperatures, we can evaluate them analytically. 

\subsubsection{Infinite temperature limit}

First consider $\beta\to0$ and expand everything to leading order in $\beta$. For instance, the susceptibility matrix becomes

\begin{equation}
    \chi = \begin{pmatrix}
    \frac{\beta}{4} & -\frac{\beta^2\mu}{4} \\
    -\frac{\beta\mu}{4} & \frac{\beta^2(J^2+\mu^2)}{4}
    \end{pmatrix}, 
\end{equation}
and hence $\gamma = \beta^3J^2/4$. In the case when we bias our chain with a temperature gradient, we can approximate 

\begin{equation}
    F_\beta^>(t) = -\frac{B_2(2Jt)}{2t}\nabla \beta+i\frac{\mu B_1(2Jt)}{4Jt}\nabla\beta,
\end{equation}
where $B_{1, 2}$ are Bessel functions of the first kind. Plugging this into Eqs.~(\ref{eq:grad_Q},~\ref{eq:grad_E_q=2}-\ref{eq:j_E_q=2}), we find 

\begin{align}
    \nabla Q &= \frac{\mu}{4}\nabla \beta, \\
    \nabla E &= -\frac{J^2}{4}\nabla \beta, \\
    j^Q &= -\frac{2\mu J_1^2}{3\pi J}\nabla \beta, \\
    j^E &= \frac{8J_1^2J}{15\pi}\nabla \beta.
\end{align}
Similarly, in the presence of a chemical potential gradient we can write 

\begin{equation}
    F_\mu^>(t) = \frac{\beta^3\mu B_2(2Jt)}{4t}\nabla \mu+i\frac{\beta B_1(2Jt)}{4Jt}\nabla\mu,
\end{equation}
and therefore deduce a new set of observables 

\begin{align}
    \nabla Q &= \frac{\beta}{4}\nabla \mu, \\
    \nabla E &= \frac{\beta^3\mu J^2}{8}\nabla \mu, \\
    j^Q &= -\frac{2\beta J_1^2}{3\pi J}\nabla \mu, \\
    j^E &= -\frac{4\beta^3\mu J_1^2J}{15\pi}\nabla \mu.
\end{align}
Together these form a set of four equations each for the diffusivity and conductivity matrices. The solutions are given by 

\begin{align}
    D &= \begin{pmatrix}
    \frac{8J_1^2}{3\pi J} & 0 \\
    -\frac{8\mu J_1^2}{15\pi J} & \frac{32J_1^2}{15\pi J}
    \end{pmatrix}, \label{eq:diffusivity_q=2_infinite}\\
    L &= \begin{pmatrix}
    \frac{2\beta J_1^2}{3\pi J} & -\frac{2\beta^2\mu J_1^2}{3\pi J} \\
    -\frac{2\beta\mu J_1^2}{3\pi J} & \frac{8\beta^2 J_1^2J}{15\pi} + \frac{2\beta^2\mu^2 J_1^2}{3\pi J}
    \end{pmatrix}, \label{eq:conductivity_q=2_infinite}
\end{align}
in agreement with the results in \figref{fig:fig4} and \figref{fig:fig5}. Note that $D_{21} = \mu(D_{22}-D_{11})$. In addition, we can check that the generalized Einstein relation $L=D\chi$ is indeed satisfied and that the thermal conductivity takes the form

\begin{equation}
    \kappa = \frac{8\beta^2 J_1^2J}{15\pi} = \frac{D_{22}\gamma}{\beta}.
\end{equation}

\subsubsection{Zero temperature limit}

Now take the opposite limit of  $\beta\to\infty$ and expand in $1/\beta$. To leading order, the susceptibility is given by 

\begin{equation}
    \chi = \begin{pmatrix}
    \frac{1}{\pi J}\sqrt{1-\left(\frac{\mu}{2J}\right)^2} & -\frac{\pi\mu}{12J^3\beta\sqrt{1-\left(\frac{\mu}{2J}\right)^2}} \\
    -\frac{\pi\mu}{12J^3\beta^2\sqrt{1-\left(\frac{\mu}{2J}\right)^2}} & \frac{\pi}{3\beta J}\sqrt{1-\left(\frac{\mu}{2J}\right)^2}
    \end{pmatrix}, 
\end{equation}
and $\gamma = \frac{\pi}{3J} \sqrt{1-\left(\frac{\mu}{2J}\right)^2}$. The non-equilibrium contribution from a temperature gradient is 

\begin{equation}
    F_\beta^>(t) = \frac{\pi e^{-i\mu t}}{3J\beta^3}\Bigg(\frac{i\mu}{4J^2\sqrt{1-\left(\frac{\mu}{2J}\right)^2}}-t\sqrt{1-\left(\frac{\mu}{2J}\right)^2}\Bigg)\nabla \beta,
\end{equation}
leading to 

\begin{align}
    \nabla Q &= \frac{\pi\mu}{12(\beta J)^3}\frac{1}{\sqrt{1-\left(\frac{\mu}{2J}\right)^2}}\nabla \beta, \\
    \nabla E &= -\frac{\pi}{3\beta^3J}\frac{1-2\left(\frac{\mu}{2J}\right)^2}{\sqrt{1-\left(\frac{\mu}{2J}\right)^2}}\nabla \beta, \\
    j^Q &= -\frac{\pi\mu J_1^2}{6\beta^3 J^4}\nabla \beta, \\
    j^E &= \frac{\pi J_1^2}{3\beta^3J^2}\left(1-3\left(\frac{\mu}{2J}\right)^2\right)\nabla \beta.
\end{align}
Analogously, the contribution due to a chemical potential gradient can be expanded as

\begin{equation}
    F_\mu^>(t) = i\frac{e^{-i\mu t}}{\pi J}\sqrt{1-\left(\frac{\mu}{2J}\right)^2}\nabla\mu,
\end{equation}
and hence our final observables are

\begin{align}
    \nabla Q &= \frac{1}{\pi J}\sqrt{1-\left(\frac{\mu}{2J}\right)^2}\nabla \mu, \\
    \nabla E &= \frac{\mu}{\pi J}\sqrt{1-\left(\frac{\mu}{2J}\right)^2}\nabla \mu, \\
    j^Q &= -\frac{J_1^2}{\pi J^2}\left(1-\left(\frac{\mu}{2J}\right)^2\right)\nabla \mu, \\
    j^E &= -\frac{\mu J_1^2}{\pi J^2}\left(1-\left(\frac{\mu}{2J}\right)^2\right)\nabla \mu.
\end{align}
Solving for the transport coefficients, we conclude that 

\begin{align}
    D &= \begin{pmatrix}
    \frac{J_1^2}{J}\sqrt{1-\left(\frac{\mu}{2J}\right)^2} & -\frac{\mu J_1^2}{4J^3\sqrt{1-\left(\frac{\mu}{2J}\right)^2}} \\
    -\frac{\pi^2\mu J_1^2}{12\beta^2J^3\sqrt{1-\left(\frac{\mu}{2J}\right)^2}} & \frac{J_1^2}{J}\sqrt{1-\left(\frac{\mu}{2J}\right)^2}
    \end{pmatrix}, \label{eq:diffusivity_q=2_zero}\\
    L &= \begin{pmatrix}
    \frac{J_1^2}{\pi J^2}\left(1-\left(\frac{\mu}{2J}\right)^2\right) & -\frac{\pi\mu J_1^2}{6\beta J^4} \\
    -\frac{\pi\mu J_1^2}{6\beta^2 J^4} & \frac{\pi J_1^2}{3\beta J^2}\left(1-\left(\frac{\mu}{2J}\right)^2\right)
    \end{pmatrix}, \label{eq:conductivity_q=2_zero}
\end{align}
which agrees with our numerics in \figref{fig:fig4} and \figref{fig:fig5}. Again, we find that $D_{21}=\mu(D_{22}-D_{11})=0$ to first order in $1/\beta$. The non-zero contribution above is a higher-order correction necessary to satisfy the Einstein relation. \eqref{eq:sigma_and_kappa} holds as expected with $\kappa = \bar{\kappa} = D_{22}\gamma/\beta$. Our answer agrees with the free-fermion calculation in Ref.~\cite{song2017} that found $\sigma=1/\pi$ and $\kappa = \pi T/3$ in units of $J\approx J_1=1$ and $\mu=0$, thus providing an independent consistency check of our methods. 

\subsection{Large \texorpdfstring{$q$}{q} limit}

We now return to the large $q$ analysis of \secref{sec:large_q_methods} and discuss the thermodynamic properties of our model in this limit. The grand canonical potential has been previously derived in Ref.~\cite{davison2017}

\begin{equation}
\begin{split}
    \Omega &= -\frac{1}{\beta}\log\left(2\cosh(\beta\mu/2)\right)\\
    &-\frac{2\pi v}{\beta q^2\cosh^2(\beta\mu/2)}\left(\tan(\frac{\pi v}{2})-\frac{\pi v}{4}\right).
    \label{eq:omega_large_q}
\end{split}
\end{equation}
To leading order in $1/q$, the charge becomes 

\begin{equation}
    Q = \frac{1}{2}\tanh(\frac{\beta\mu}{2}) + \mathcal{O}\left(\frac{1}{q^2}\right).
    \label{eq:charge_large_q}
\end{equation}
It follows that at low temperatures, the chemical potential should scale as $\mu\sim T$ to maintain a constant charge. This stems from a failure of the infinite $q$ and infinite $\beta$ limits to commute, which is an inherent shortcoming of this expansion~\cite{Bhattacharya2017}. 

Taking a second derivative, we find the susceptibilities

\begin{align}
    \chi_{11} &= \frac{\beta}{4\cosh^2(\beta\mu/2)} + \mathcal{O}\left(\frac{1}{q^2}\right),\\
    \chi_{12} &= -\frac{\beta^2\mu}{4\cosh^2(\beta\mu/2)} + \mathcal{O}\left(\frac{1}{q^2}\right), \\
    \chi_{22} &= \frac{(\beta\mu)^2}{4\cosh^2(\beta\mu/2)}\left( 1+\left(\frac{2\pi v}{q\beta\mu}\right)^2\frac{1}{\frac{\pi v}{2}\tan(\frac{\pi v}{2})+1}\right),
\end{align}
where the second term in $\chi_{22}$ is necessary to obtain the leading order contribution to 

\begin{equation}
    \gamma = \frac{\beta}{q^2\cosh^2(\beta\mu/2)}\frac{(\pi v)^2}{\frac{\pi v}{2}\tan\left(\frac{\pi v}{2}\right)+1}. 
\end{equation}
The prefactors can also be written in terms of the charge $\cosh^2(\beta\mu/2)=(1-4Q^2)^{-1}$. Note that in the zero-temperature limit, at fixed charge, the compressibility $\chi_{11}$ diverges with $\beta$, which is unphysical. The correct behavior can be recovered by keeping the next order term in the large $q$ expansion and taking the temperature to zero first~\cite{davison2017}. This is another example where the order in which we take the limits matters. 

In the case of transport, we first imagine maintaining a constant chemical potential gradient across the chain held at a fixed temperature. The non-equilibrium contribution defined in \eqref{eq:def_f} can be computed via the chain rule 

\begin{equation}
\begin{split}
    f_\mu(t) &= -(q-2)\frac{\beta^2\J}{4}\tanh(\frac{\beta\mu}{2})\frac{\cos(\frac{\pi v}{2})}{\frac{\pi v}{2}\tan(\frac{\pi v}{2})+1}\\
    &\cdot \left(\left(1-\frac{2it}{\beta}\right)\tan(\frac{\pi v}{2}-\frac{i\pi vt}{\beta})-\tan(\frac{\pi v}{2})\right).
\end{split}
\end{equation}
The energy gradient is obtained by direct integration 

\begin{equation}
    \nabla E = \frac{\beta\J\tanh(\frac{\beta\mu}{2})}{2q\cosh^2(\frac{\beta\mu}{2})}\Bigg(\frac{\frac{\pi v}{2}\cos(\frac{\pi v}{2})}{\frac{\pi v}{2}\tan(\frac{\pi v}{2})+1}+\sin\left(\frac{\pi v}{2}\right)\Bigg)\nabla \mu. 
\end{equation}
However, the energy current is a bit more subtle. It turns out that the overall contribution from the terms proportional $f_\mu$ is zero, so we only have to consider the last term in \eqref{eq:j_++_large_q}. We use the identity 

\begin{equation}
    \int_{0}^{\infty}\diff t \int_{t}^{\infty} \diff t' \im\left[e^{g(t)+g(t')}\right] = -\frac{\sin(\pi v)}{2\J^2},
\end{equation}
and finally arrive at 

\begin{equation}
    j^E = -\frac{\beta\J_1^2\tanh(\frac{\beta\mu}{2})}{4q\cosh^2(\frac{\beta\mu}{2})}\sin(\pi v)\nabla \mu. 
\end{equation}
The other two observables related to charge transport are given by Eqs.~(\ref{eq:grad_Q_large_q},~\ref{eq:j_Q_large_q}).

Next, we will consider a slightly simpler setup where we impose both a temperature and chemical potential gradient, but maintain a constant charge $\nabla Q = 0$. From \eqref{eq:charge_large_q} this is equivalent to holding $\beta\mu$ constant along the chain and setting $\beta\nabla\mu = -\mu\nabla\beta$. In linear response, this corresponds to an additive contribution from both $f_\mu$ and $f_\beta$, which we denote by 

\begin{equation}
    f_Q(t) = \frac{(1+i\J t\sin(\frac{\pi v}{2}))\tan(\frac{\pi v}{2}-\frac{i\pi vt}{\beta})-\tan(\frac{\pi v}{2})}{\frac{\beta}{\pi v}\left(\frac{\pi v}{2}\tan(\frac{\pi v}{2})+1\right)}\nabla \beta
\end{equation}
This is exactly the answer we found for a Majorana chain~\cite{zanoci2022}, up to a constant prefactor. Therefore, we can follow the same calculations to find 

\begin{align}
    \nabla E &= -\frac{\J}{q^2\beta\cosh^2(\beta\mu/2)}\frac{\pi v\cos(\frac{\pi v}{2})}{\frac{\pi v}{2}\tan(\frac{\pi v}{2})+1}\nabla \beta, \\
    j^E &= \frac{\J_1^2}{q^2\beta\cosh^2(\beta\mu/2)}\frac{\pi v}{3}\cos^2\left(\frac{\pi v}{2}\right) \nabla \beta.
\end{align}
Note that the charge current vanishes as expected, since $j^Q\sim(C_\mu+C_\beta)=0$ for this setup. 

Finally, combining all the results into a system of equations, we deduce the diagonal diffusivity entries

\begin{align}
    D_{11} &= \frac{\J_1^2}{\J}\cos(\frac{\pi v}{2}),\\
    D_{22} &= \frac{\J_1^2}{3\J}\left(\frac{\pi v}{2}\sin(\frac{\pi v}{2})+\cos(\frac{\pi v}{2})\right),
\end{align}
and the off-diagonal values $D_{12} = 0$ and $D_{21} = \mu(D_{22}-D_{11})$. Similarly, the conductivities are 

\begin{align}
    \sigma &= \frac{\J_1^2}{\J^2}\frac{\pi v}{4\cosh^2(\beta\mu/2)}, \label{eq:sigma_large_q}\\
    \alpha &=-\beta\mu \sigma =  -\beta\mu\frac{\J_1^2}{\J^2}\frac{\pi v}{4\cosh^2(\beta\mu/2)}, \\
    \bar{\kappa} &= \beta\mu^2\frac{\J_1^2}{\J^2}\frac{\pi v}{4\cosh^2(\beta\mu/2)}\left( 1+\frac{4}{3}\left(\frac{\pi v}{q\beta\mu}\right)^2\right),
\end{align}
and we can check the dependence in \eqref{eq:sigma_and_kappa}

\begin{equation}
    \kappa = \frac{\J_1^2}{\J^2}\frac{(\pi v)^3}{3\beta q^2 \cosh^2(\beta\mu/2)} = \frac{D_{22}\gamma}{\beta}. 
    \label{eq:kappa_large_q}
\end{equation}

The large $q$ approximation has given us remarkably simple closed-form answers for all the transport coefficients. The generalized Einstein relation now can be checked explicitly. Notice that the conductivities, just like the susceptibilites, are suppressed by a factor of $\cosh^2(\beta\mu/2)$. This again suggests that we should scale our parameters to maintain a finite value of $\beta\mu$. In our numerics at finite $q$, we maintain both $J$ and $\mu$ constant while sweeping a wide range of temperatures. Therefore, in order to make the results of the large $q$ approximation consistent in this regime, we will restrict ourselves to the case $\mu=0$ when comparing to fixed $q$ results. 

\section{Thermopower}
\label{sec:appendixD}

\begin{figure}[tp]
	\begin{center}
	\includegraphics[width = \columnwidth]{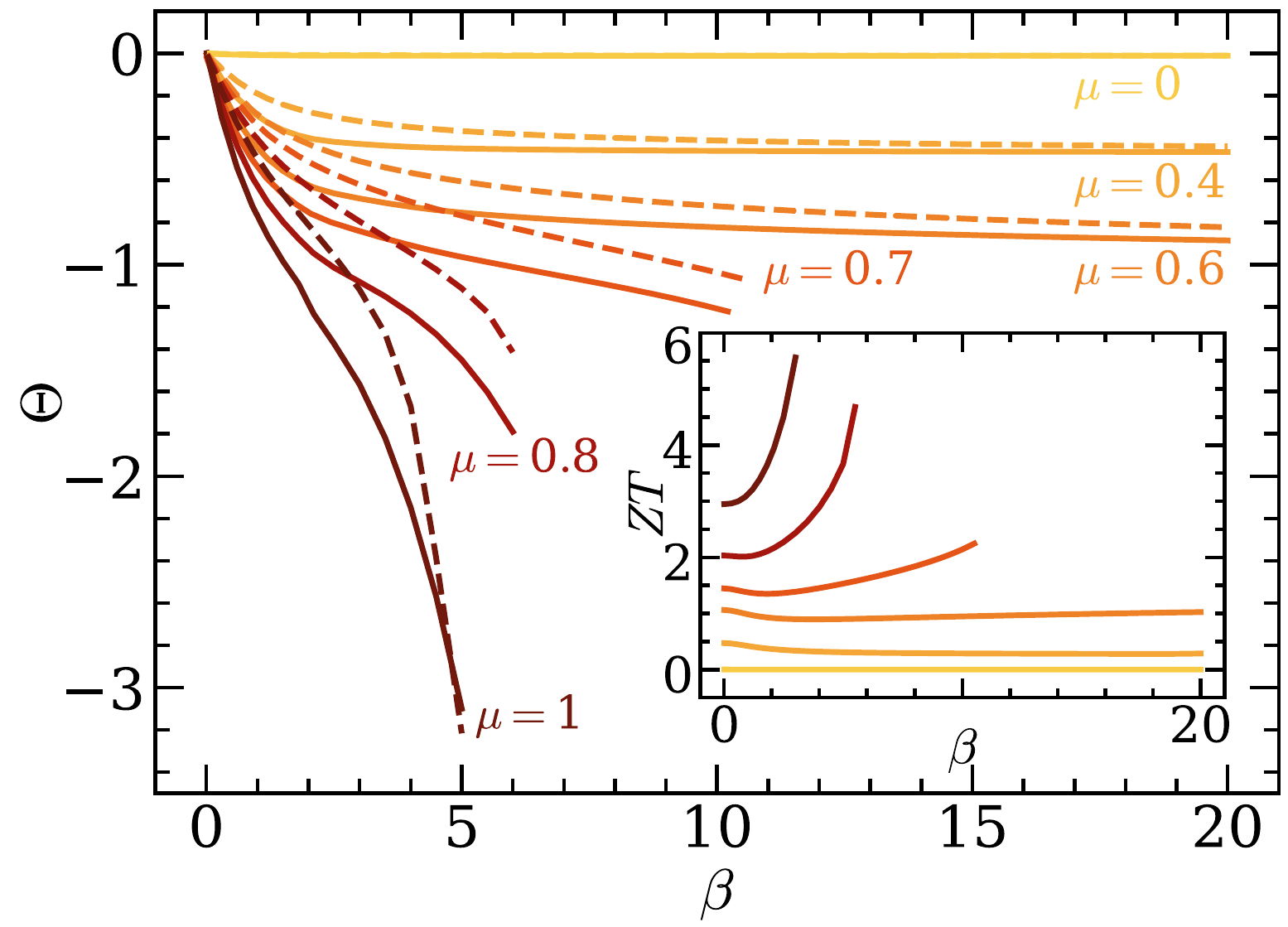}
	\caption{Temperature dependence of the thermopower (solid lines) for the SYK model with $q = 4$. At low temperatures in the SYK-like phase, $\Theta$ approaches the conformal answer $2\pi\mathcal{E}$ (dashed lines). The inset shows the thermoelectric figure of merit $ZT$ for the same parameters.}
	\label{fig:fig10}
	\end{center}
\end{figure}

If a temperature gradient is applied to across a material with free charge carriers, a potential gradient will arise as a result of the carriers' motion from hot to cold areas. The magnitude of this thermoelectric effect is characterized by the thermopower $\Theta$, also known as the Seebeck coefficient. The thermopower is defined as the ratio of the induced potential gradient $\nabla \mu$ to the applied temperature gradient $\nabla T$ after the system has reached a steady state with no charge current~\cite{kubo1985statistical,de1984non,forster1975hydrodynamic}. \eqref{eq:conductivity} implies 

\begin{equation}
    \Theta = -\frac{\nabla \mu}{\nabla T} = \frac{\alpha}{\sigma}. 
\end{equation}
Materials with high thermopower are very important for building efficient thermoelectric generators and coolers. A useful metric for quantifying the effectiveness of a thermoelectric material for practical applications is the dimensionless thermoelectric figure of merit 

\begin{equation}
    ZT = \frac{\sigma\Theta^2T}{\kappa} = \frac{\alpha^2T}{\sigma\kappa} = \frac{\bar{\kappa}}{\kappa} - 1.
\end{equation}
For conventional metals and insulators, $ZT$ is at most of order one~\cite{skinner2018}. Recently, it has been shown that the thermopower of Dirac and Weyl semimetals in an external magnetic field grows linearly with the magnitude of the field and can reach extremely high values~\cite{skinner2018,kozii2019,Han2020}. This makes it possible to have $ZT\gg1$ in these materials.  

We now calculate the thermopower and thermoelectric figure of merit for the complex SYK model using our previous results for the conductivity matrix. We showcase our findings for $q=4$ in \figref{fig:fig10}. The thermopower reaches a non-zero constant at low temperatures and increases with $\mu$ as we approach the phase transition. The thermoelectric figure of merit is of order one and has a similar dependence on $\beta$ and $\mu$.  This is consistent with a previous analysis of thermopower in SYK models and holographic theories~\cite{davison2017,kruchkov2020}, where the authors showed that in the low temperature limit

\begin{equation}
    \Theta = 2\pi\mathcal{E},
\end{equation}
where $\mathcal{E}$ is the particle-hole asymmetry of the fermionic spectral function~\cite{georges2001,Gu2020,davison2017}, which can be expressed in terms of charge via a Luttinger-Ward identity

\begin{align}
    e^{2\pi\mathcal{E}} &= \frac{\sin(\pi/q+\theta)}{\sin(\pi/q-\theta)},\\
    Q &= -\frac{\theta}{\pi} - \left(\frac{1}{2}-\frac{1}{q}\right)\frac{\sin(2\theta)}{\sin(2\pi/q)}.
\end{align}
Note that with our conventions, both $\mathcal{E}$ and $\theta$ are negative. By combining this result with the Wiedemann-Franz ratio in \eqref{eq:WF_ratio}, we find that the figure of merit should scale as $ZT = 3q^2\mathcal{E}^2$ at low temperatures. The asymmetry diverges as the charge becomes larger when leaving the SYK-like phase, which explains the scaling in \figref{fig:fig10}. 

The non-vanishing thermopower in the limit of zero temperature is a remarkable feature of the SYK model. It can be related to the existence of a finite zero-temperature entropy $S_0$ via an exact Kelvin formula~\cite{georges2001,davison2017}. Recently, it has been suggested that thermopower measurements can serve as a direct probe of the residual entropy $S_0$~\cite{kruchkov2020}. This low-temperature behavior is in stark contrast to that of a Fermi liquid, whose thermopower vanishes linearly with $T$. In fact, we can see this explicitly for the $q=2$ SYK, where \eqref{eq:conductivity_q=2_zero} implies $\Theta\sim T$ and $ZT\sim T^2$. 

The thermoelectric parameters can be computed exactly in the large $q$ limit. We find the simple expressions $\Theta = -\beta\mu$ and $ZT = \frac{3q^2(\beta\mu)^2}{4(\pi v)^2}$. The thermopower is directly related to $Q$ through \eqref{eq:charge_large_q} and is constant at fixed charge. At high temperatures, $\pi v\approx \beta\J$ and the figure of merit remains finite with $ZT\sim q^2\mu^2$. This precisely matches our results in the inset of \figref{fig:fig10}. Given this scaling of $ZT$, it seems possible to further increase the thermodynamic figure of merit by going to higher $q$ SYK models.

\end{document}